\documentclass[twocolumn,trackchanges]{aastex631}
\usepackage{natbib, aas_macros}
\usepackage{threeparttable}
\begin{document}

\title{Multi-line assessment of narrow-line regions in $z \sim$ 3 radio galaxies}

\author{Koki Terao}
\affiliation{Astronomical Institute, Tohoku University, Aramaki, Aoba-ku, Sendai, Miyagi 980-8578, Japan}

\author{Tohru Nagao}
\affiliation{Research Center for Space and Cosmic Evolution, Ehime
  University, Bunkyo-cho 2-5, Matsuyama, Ehime 790-8577, Japan}

\author{Kyoko Onishi}
\affiliation{Research Center for Space and Cosmic Evolution, Ehime
  University, Bunkyo-cho 2-5, Matsuyama, Ehime 790-8577, Japan}
\affiliation{Department of Space, Earth and Environment, Chalmers University of Technology,
  Onsala Space Observatory, SE-439 92 Onsala, Sweden}

\author{Kenta Matsuoka}
\affiliation{Utena Meishu Company, Limited, 519-1 Morimatsu-machi, Matsuyama, Ehime 791-1113, Japan}

\author{Masayuki Akiyama}
\affiliation{Astronomical Institute, Tohoku University, Aramaki, Aoba-ku, Sendai, Miyagi 980-8578, Japan}

\author{Yoshiki Matsuoka}
\affiliation{Research Center for Space and Cosmic Evolution, Ehime
  University, Bunkyo-cho 2-5, Matsuyama, Ehime 790-8577, Japan}

\author{Takuji Yamashita}
\affiliation{Research Center for Space and Cosmic Evolution, Ehime
  University, Bunkyo-cho 2-5, Matsuyama, Ehime 790-8577, Japan}
\affiliation{National Astronomical Observatory of Japan, 2-21-1 Osawa, Mitaka, Tokyo 181-8588, Japan}

%%%%%%%%%%%%%%%%%%%
\begin{abstract}
  In this paper, we utilize high-quality rest-UV spectra of three radio galaxies at $z \sim$ 3 observed with VLT/FORS2 to measure the flux of several emission lines including relatively faint ones, such as N~{\sc iv}]$\lambda$1486, O~{\sc iii}]$\lambda$1663, and [Ne~{\sc iv}]$\lambda$2424. Additionally, we collect fluxes of faint rest-UV emission lines in 12 $z \sim$ 3 radio galaxies from the literature.
      Previously, physical and chemical properties of narrow-line regions (NLRs) in high-$z$ active galactic nuclei (AGNs) have been investigated mostly by using only strong rest-UV emission-lines (e.g., N~{\sc v}$\lambda$1240, C~{\sc iv}$\lambda$1549, He~{\sc ii}$\lambda$1640, and C~{\sc iii}]$\lambda$1909).
      Such strong-line diagnostics are based on various assumptions due to the limitation in the number of available emission-line constraints.
   In this work, both physical and chemical properties of NLR clouds in each object are estimated by fitting detailed photoionization models to the measured emission-line fluxes.
   We confirm that the metallicity of NLRs in AGNs at $z \sim$ 3 is solar or super-solar, without assuming the gas density and ionization parameter thanks to the constrains from the faint emission lines.
  This result suggests that high-$z$ radio galaxies are already chemically matured at $z \sim$ 3.
\end{abstract}


%%%%%%%%%%%%%%%%%%%%%%%%%%%%%%%%%%%%%%%%%%%%%%%%%%%%%%%%%%%%%%%%%%%%%%%%%%%%%%%
\section{Introduction}
Understanding the formation and evolution of galaxies is one of hot topics in modern astronomy.
For tackling this topic, investigating the nature of components of galaxies such as stars, gas, and dark matter at various cosmic epochs is a fundamental approach. 
Especially, it is important to understand the redshift evolution of interstellar medium (ISM) properties that are characterized by physical and chemical parameters such as gas density, ionization parameter, and metallicity.
In particular, metallicity reflects star-formation and gas inflow/outflow history of galaxies \citep[e.g.,][]{2008ApJ...674..151E,2013ApJ...772..119L,2015ApJ...808..129L,2016MNRAS.455.1218B,2018MNRAS.477...56V}.
ISM properties in low-redshift galaxies are usually estimated with optical emission-line diagnostics \citep[e.g.,][]{2006A&A...448..955I,2006A&A...459...85N,2008ApJ...681.1183K,2017MNRAS.465.1384C,2020MNRAS.491..944C}.
For galaxies at high redshifts ($z >$ 1), the rest-frame optical emission lines are shifted into near-infrared and thus measurements of their strength are more challenging.
Recently, the ISM properties of such high-$z$ star-forming galaxies have been also investigated \citep[e.g.,][]{2009ApJ...706.1364F,2012PASJ...64...60Y,2013ApJ...763....9Y,2014ApJ...792....3M,2014ApJ...781...21N,2018MNRAS.477.2098N,2020MNRAS.491.1427S} and the observed evolution of the emission line properties are consistently explained with photoionization models with extreme ISM conditions \citep[e.g.,][]{2013ApJ...774..100K}.
However, the determination of detailed physical parameters of ISM for high-$z$ star-forming galaxies is generally difficult because their emission lines are faint.

On the other hand, active galactic nuclei (AGNs) are luminous and thus their emission-line fluxes can be measured even in high-$z$ universe.
One important advantage of the spectroscopic study for AGNs in comparison to that for star-forming galaxies is that AGNs show strong rest-frame UV emission lines, which are faint in star-forming galaxies.
Many spectroscopic observations for high-$z$ AGNs have been carried out for various diagnostic studies \citep[e.g.,][]{1999A&A...351...47V,2000A&A...362..519D,2001A&A...366....7V,2004MNRAS.351..997S,2006A&A...447..863N,2008MNRAS.383...11H,2011A&A...532L..10M,2016MNRAS.456.3354F}. 
Particularly, spectroscopic properties of narrow-line regions (NLRs) in AGNs have been often investigated because NLR gas clouds distribute upto $\sim$kpc scale that can trace the far larger scale than broad-line regions (BLRs; located at $\lesssim$1 pc typically) and thus the NLR is more appropriate to study the ISM property of AGN host galaxies.
Since NLR clouds are mostly ionized through the photoionization process \citep[e.g.,][]{1996A&A...312..365B,1997A&A...323...31K,2004ApJS..153...75G,2016ApJ...833..266T}, ISM properties of host galaxies can be studied through detailed comparisons between photoionization models and emission-line spectra of NLRs. 

For studying spectroscopic properties of NLRs in AGNs, more reliable measurements on the emission line properties of the NLRs can be obtained for type-2 AGNs than for type-1 AGNs.
This is because the strong broad-line emission from BLRs in type-2 AGNs is blocked by optically-thick dusty tori, and thus both of forbidden and permitted emission lines can be used for various diagnostic studies without being affected by the BLR emission, which only appears in the permitted emission lines.
Among some populations of type-2 AGNs, high-$z$ radio galaxies (HzRGs) have often been targeted in high-$z$ spectroscopic studies since they are easier to be found compared to radio-quiet ones.
In previous spectroscopic studies of HzRGs, strong UV lines (such as N~{\sc v}$\lambda$1240, C~{\sc iv}$\lambda$1549, He~{\sc ii}$\lambda$1640, and C~{\sc iii}]$\lambda$1909) have been used to characterize their ISM properties \citep[e.g.,][]{2006A&A...447..863N,2007MNRAS.375.1299V,2008MNRAS.383...11H,2009A&A...503..721M,2014MNRAS.443.1291D,2017MNRAS.465.2698M,2018A&A...616L...4M}.
The results of these studies suggest that the HzRGs had already chemically matured by $z \sim$ 4, and no significant redshift evolution in 1 $< z < $ 4 is observed.

In the previous studies, however, emission-line diagnostic studies have required assumptions on some ISM parameters; in other words, it has been difficult to determine the physical and chemical parameters of the ISM simultaneously due to the small number of detected emission lines.
For example, the gas metallicity has been sometimes estimated with C~{\sc iv}/He~{\sc ii} and C~{\sc iii}]/C~{\sc iv} flux ratios assuming a fixed gas density \citep[e.g.,][]{2009A&A...503..721M}. The C~{\sc iv}/He~{\sc ii} flux ratio, however, depends on the gas density of ionizing clouds.
  This flux ratio can vary in $\sim$ 2 dex when log $n$ (in cm$^{-3}$) changes from $\sim$ 2 to $\sim$ 6 \citep{2009A&A...503..721M}.
  Consequently, the results of the metallicity estimations that do not take into account the dependence of those flux ratios on gas density can vary by factor 2 depending on the assumed gas density \cite[e.g.,][]{2018A&A...616L...4M}.
However, such degeneracies can be solved by utilizing not only traditionally-used strong emission lines \citep{2018MNRAS.474.3649S} but also faint emission lines.
Therefore, in this paper, we utilize high signal-to-noise (SN) rest-UV spectra of HzRGs to measure the flux of several emission lines including relatively weak ones.
We investigate the physical and chemical properties of the ISM in HzRGs without assuming the gas density and ionization parameter.

This paper is structured as follows. In Section 2, we describe the spectroscopic data and reduction processes. We show the results of the data analysis in Section 3 and describe our photoionization models in Section 4. We discuss the interpretation of our results in Section 5 and draw the conclusion of this paper in Section 6. Throughout this paper, we assume $\Omega_{\rm M}=$0.3, $\Omega_{\rm \Lambda}$=0.7, and $H_{0}=$70 km s$^{-1}$ Mpc$^{-1}$.

%%%%%%%%%%%%%%%%%%%%%%%%%%%%%%%%%%%%%%%%%%%%%%%%%%%%%%%%%%%%%%%%%%%%%%%%%%%%%%%%%%%%%%
\section{Data}
\subsection{Our targets and data reduction}
For analyzing narrow emission lines including faint ones, we investigate the high-quality rest-frame UV spectra of HzRGs presented by \citet{2009A&A...503..721M}. Among the nine HzRGs studied by \citet{2009A&A...503..721M}, we specifically focus on three HzRGs (TN J0920$-$0712, 4C 24.28, and USS 1545$-$234) whose high-SN spectra show at least six emission lines with SN $>$ 5.
Thanks to the high-SN, we detect N~{\sc iv}]$\lambda$1486, O~{\sc iii}]$\lambda$1663, and [Ne~{\sc iv}]$\lambda$2424 emission lines, which are weaker than C~{\sc iv}, He~{\sc ii}, and C~{\sc iii}] emission lines.
HzRGs in \citet{2009A&A...503..721M} were originally selected from a HzRG catalog \citep{2000A&A...362..519D} with the following criteria: 1). redshift being higher than 2.7; 2). emission line flux of C~{\sc iv}, He~{\sc ii}, and C~{\sc iii}] had not been measured.
  Basic properties of the three objects are summarized in Table~\ref{tab:targets}.
  The observations were carried out using FORS2 (FOcal Reducer and low dispersion Spectrograph 2; \citealt{1998Msngr..94....1A}) at VLT (Very Large Telescope) between 2005 October and 2006 October (PI: Tohru Nagao). The slit width was 1$^{\prime\prime}$. The spectral resolution was $R \sim$ 500, which was measured using the width of sky emission lines. The typical seeing during the observations was $\sim$ 1$^{\prime\prime}$.4 which is broader than the slit width. The details of the observations are described in \citet{2009A&A...503..721M}.

\begin{table*}[htb]
  \centering
  \caption{Target properties}
  \begin{tabular}{lcccl}\hline
      Name & $z^a$ & $E(B-V)^b$ & Exp. (min) & Date of observations \\ \hline
      TN J0920$-$0712 & 2.758 & 0.041 & 180 & 2006 Apr. 3, 4\\
      4C 24.28 & 2.913 & 0.018 & 180 & 2006 Apr. 23\\
      USS 1545$-$234 & 2.751 & 0.257 & 240 & 2006 Apr. 5, 23, 24\\
      \hline
  \end{tabular}
  \label{tab:targets}
  \tablecomments{$^a$Redshifts are determined from the observed C~{\sc iv} wavelength. \\ $^b$Galactic extinction \citep{1998ApJ...500..525S}.}
\end{table*}

In this work, we are specifically interested in relatively faint emission-lines of the spectra in order to increase available emission lines for estimating ISM parameters.
However, \citet{2009A&A...503..721M} did not correct the effect of the atmospheric absorption because they focused only on strong emission lines whose wavelengths were not affected by the atmospheric absorption.
Thus we reanalyzed the spectra with additional procedure to correct the atmospheric absorption which is not negligible for faint emission lines.
The data reduction of the three HzRGs is briefly described below, which is basically the same as described in \citet{2009A&A...503..721M} except for some additional procedures.
The data analysis was performed with the IRAF software \citep{1986SPIE..627..733T,1993ASPC...52..173T}.
We adopt a usual manner to analyze the data i.e., bias subtraction using average bias frames, flat fielding, cosmic ray subtraction, wavelength calibration using sky emission-lines, sky subtraction, spectral extraction from two-dimensional spectra by adopting 2$^{\prime\prime}$.25 (9 pixels) aperture, and flux calibration by standard stars. The Galactic reddening maps from \citet{1998ApJ...500..525S} and extinction law from \citet{1989ApJ...345..245C} were adopted for the correction of the Galactic reddening of the three targets.
Sky subtractions were performed using averaged sky spectrum created by the region free from the target light in the observed 2d spectrum.
Cosmic-ray events were removed using the \texttt{lacos\_spec} task \citep{2001PASP..113.1420V} instead of the \texttt{fixpix} task used in \citet{2009A&A...503..721M}.
The spectra of targets were divided by the reduced spectra of standard stars to correct for atmospheric absorption features, which were not corrected in the previous analysis.

In general, emission lines from NLRs in AGNs are affected by the internal dust reddening caused in their host galaxies, not only in our Galaxy \citep[e.g.,][]{2003ApJ...583..159H,2006MNRAS.366..480G,2012MNRAS.427.1266V,2016MNRAS.461.4227H,2017ApJ...846..102M,2019MNRAS.483.1722L}.
In order to correct for the internal reddening, the Balmer decrement (H$\alpha$/H$\beta$ flux ratio) is usually evaluated.
However, it is not easy to estimate the amount of the internal reddening for HzRGs in our sample, since Balmer lines shift into near-infrared wavelength that were not measured for our sample.
In this work, we did not correct for such internal extinction based on the following consideration.
For low-redshift radio galaxies, \citet{1987MNRAS.227...97R} investigated the Balmer decrement of 11 radio galaxies at $z < 0.1$ and showed that extended emission-line regions of more than half of their targets show the flux ratio consistent with the case B, thus their internal dust reddening is negligible.
For HzRGs at 1.4 $< z <$ 2.6, \citet{2008MNRAS.383...11H} investigated flux ratios of H$\alpha$/H$\beta$ and/or He~{\sc ii}$\lambda$1640/He~{\sc ii}$\lambda$4686 of 11 HzRGs, and reported that more than half of them show a small extinction ($A_V < 0.5$ mag).
Therefore we assume that the internal dust extinction is negligible for our HzRG sample.
We will evaluate how extinction correction could affect our results in Section 5.1.

%%%%%%%%%%%%%%%%%%%%%%%%%%%%%%%%%%%%%%%%%%%%
\subsection{Additional rest-UV data from the literature}
In addition to the re-analyzed data described in Section 2.1, we collect rest-UV emission-line fluxes of 12 HzRGs that show various rest-UV emission lines thanks to high-SN spectra, including relatively weak ones such as N~{\sc iv}]$\lambda$1486, O~{\sc iii}]$\lambda$1663, and [Ne~{\sc iv}]$\lambda$2424.
The emission-line data of 10 $z \sim$ 3 HzRGs were obtained from \citet{2001A&A...366....7V} and \citet{2008MNRAS.383...11H}. These targets were selected from the ultra-steep spectrum (USS) radio galaxy survey \citep{1995A&AS..114...51R} for $z >$ 2.2 and $R \sim 21 - 23$ mag. These spectra were observed with LRISp (the spectro-polarimetric mode of Low-Resolution Imaging Spectrometer; \citealt{1995PASP..107..179G,1995PASP..107..375O}) on the Keck II telescope.
Another spectrum of a $z \sim$ 2 HzRG (NVSS J002402$-$325253) obtained with the FORS1 on VLT is collected from \citet{2006MNRAS.366...58D}. This target had been selected from the USS radio galaxy sample \citep{2004MNRAS.347..837D}. 
The other spectrum is 3C 256 ($z =1.824$) which was observed with the double spectrograph \citep{1982PASP...94..586O} on the Hale telescope \citep{1999ApJ...525..659S}.

%%%%%%%%%%%%%%%%%%%%%%%%%%%%%%%%%%%%%%%%%%%%%%%%%%%%%%%%%%%%%%%%%%%%%%%%%
\section{Results}
\subsection{Rest-frame UV spectra of the three HzRGs}
\begin{figure}[htb]
  \epsscale{1.3}
  \plotone{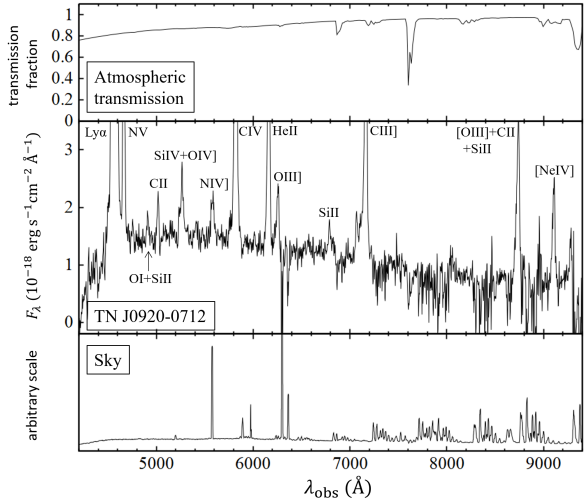}
  \caption{Rest-frame UV spectrum of TN J0920$-$0712 (middle panel). The detected emission lines are labeled. Top panel shows a typical atmospheric transmission at the VLT site \citep{2012A&A...543A..92N,2013A&A...560A..91J}. Bottom panel shows a typical sky spectrum obtained during our runs.}
  \label{fig:j0920}
\end{figure}

\begin{figure}[htb]
  \epsscale{1.3}
  \plotone{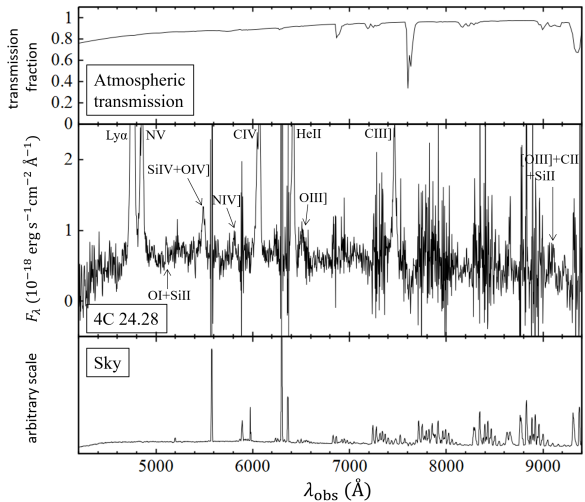}
  \caption{Same as Figure~\ref{fig:j0920} but for 4C 24.28.}
  \label{fig:4c2428}
\end{figure}

\begin{figure}[htb]
  \epsscale{1.3}
  \plotone{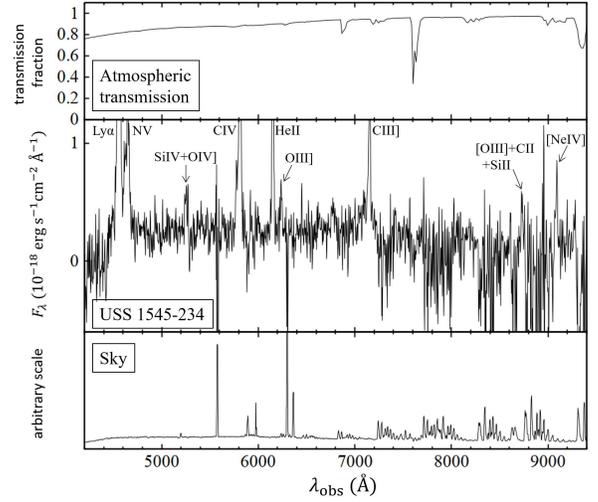}
  \caption{Same as Figure~\ref{fig:j0920} but for USS 1545$-$234.}
  \label{fig:uss1545}
\end{figure}

The newly reduced spectra of the three objects described in Section 2.1 are shown in Figures \ref{fig:j0920}--\ref{fig:uss1545}. In these figures, we also show typical atmospheric transmission and sky spectrum\footnote
{This sky spectrum was obtained from the SkyCalc sky model calculator \citep{2012A&A...543A..92N,2013A&A...560A..91J} provided by ESO (https://www.eso.org/observing/etc/bin/gen/form?INS.MODE=\\swspectr+INS.NAME=SKYCALC).}.
As shown in the figures, there are weak absorption lines and air-grow emission lines around the detected weak emission lines, while these effects are properly corrected.
From the spectra, 13 emission-lines were detected with SN$>$3 in TN J0920$-$0712. The spectra of 4C 24.28 and USS 1545$-$234 show 10 and 9 emission-lines, respectively (Table \ref{tab:line_flux}).

The fluxes, central wavelengths, FWHMs, and observed equivalent widths (EWs) of the detected emission lines with S/N $>$ 3 were measured with the IRAF task {\tt splot} assuming a single Gaussian profile.
Here the flux errors given in Table~\ref{tab:line_flux} include the uncertainty due to the pixel-to-pixel variance and the estimation of the continuum level.
Note that narrow emission lines in AGN spectra have been sometimes fitted with more sophisticated methods such as multi-component Gaussian profile \citep[e.g.,][]{1991ApJ...369..331V,2005ApJ...627..721G,2013MNRAS.433..622M}.
However, such profiles require a larger number of free parameters than the single Gaussian profile.
Since the strength of the weak emission lines can be described by the single Gaussian profile, we adopt the single Gaussian profile.
The emission-line properties obtained through the fit are given in Table~\ref{tab:line_flux}.
To confirm the consistency between the emission-line fluxes given in \citet{2009A&A...503..721M} and our results, we compare the emission-line flux ratios obtained in this work and in \citet{2009A&A...503..721M}.
The C~{\sc iii}]/C~{\sc iv} flux ratio of TN J0920$-$0712 is 0.615$\pm$0.030 in this work and 0.578$\pm$0.018 in \citet{2009A&A...503..721M}, thus these values are consistent within the 1$\sigma$ error range. The remaining two objects also show consistent emission-line flux ratios with \citet{2009A&A...503..721M}; the C~{\sc iii}]/C~{\sc iv} flux ratio of 4C 24.28 is 0.657$\pm$0.035 in \citet{2009A&A...503..721M} and 0.719$\pm$0.056 in this work, and that of USS 1545$-$234 is 0.451$\pm$0.024 in \citet{2009A&A...503..721M} and 0.426$\pm$0.036 in this work.

\begin{table*}[htb]
  \centering
  \caption{Detected emission lines}  
  \begin{tabular}{llccccc}\hline\hline
    Name & line & flux & $\lambda_{\rm obs}^a$ & FWHM$_{\rm obs}^b$ & FWHM$_{\rm corr}^c$& EW$_{\rm obs}$\\
    & & (10$^{-17}$ erg s$^{-1}$ cm$^{-2}$) & (\AA) & (\AA) & (km s$^{-1}$) & (\AA)\\\hline
    TN J0920$-$0712
    &Ly$\alpha\lambda$1216                              & 305$\pm$1     & 4568.34$\pm$0.01 & 30.09$\pm$0.05 & 1881$\pm$3   & 3179$\pm$341\\
    &N~{\sc v}$\lambda$1240                             & 7.14$\pm$0.51 & 4664.63$\pm$0.13 & 26.91$\pm$0.83 & 1662$\pm$57  & 36.88$\pm$3.81\\
    & O~{\sc i}+Si~{\sc ii}$\lambda$1305                & 1.15$\pm$0.17 & 4912.57$\pm$0.91 & 31.46$\pm$5.20 & 1821$\pm$332 & 11.18$\pm$1.37\\
    & C~{\sc ii}$\lambda$1335                           & 1.62$\pm$0.16 & 5017.76$\pm$0.29 & 19.56$\pm$1.44 & 1001$\pm$101 & 12.80$\pm$1.87\\
    & Si~{\sc iv}$\lambda$1397+O~{\sc iv}]$\lambda$1402 & 4.24$\pm$0.41 & 5265.02$\pm$0.17 & 38.32$\pm$1.13 & 2098$\pm$67  & 26.52$\pm$1.24\\
    & N~{\sc iv}]$\lambda$1486                          & 2.50$\pm$0.19 & 5581.59$\pm$0.10 & 37.68$\pm$0.94 & 1933$\pm$53  & 19.85$\pm$0.82\\
    & C~{\sc iv}$\lambda$1549                           & 29.6$\pm$0.8  & 5821.40$\pm$0.01 & 26.34$\pm$0.09 & 1217$\pm$5   & 216.6$\pm$4.6\\
    &He~{\sc ii}$\lambda$1640                           & 18.3$\pm$1.0  & 6161.46$\pm$0.01 & 21.24$\pm$0.16 & 841$\pm$10   & 144.6$\pm$5.6 \\
    &O~{\sc iii}]$\lambda$1663                          & 3.32$\pm$0.48 & 6254.25$\pm$0.32 & 30.40$\pm$0.89 & 1328$\pm$47  & 28.48$\pm$1.58\\
    &Si~{\sc ii}$\lambda$1814                           & 2.77$\pm$0.31 & 6793.44$\pm$0.90 & 61.13$\pm$6.67 & 2629$\pm$302 & 24.50$\pm$3.14\\
    &C~{\sc iii}]$\lambda$1909                          & 18.2$\pm$1.1  & 7163.06$\pm$0.01 & 29.62$\pm$0.28 & 1085$\pm$14  & 200.1$\pm$8.0\\
    &[O~{\sc iii}]+C~{\sc ii}+Si~{\sc ii}$\lambda$2322  & 11.1$\pm$0.83 & 8735.29$\pm$0.12 & 38.19$\pm$1.03 & 1165$\pm$40  & 216.5$\pm$25.6\\
    &[Ne~{\sc iv}]$\lambda$2424                         & 6.18$\pm$0.50 & 9104.60$\pm$0.09 & 31.24$\pm$0.44 & 836$\pm$18   & 87.90$\pm$3.73\\
    \hline
    4C 24.28
    &Ly$\alpha\lambda$1216                              & 46.5$\pm$0.4  & 4761.77$\pm$0.03 & 29.93$\pm$0.20 & 1786$\pm$13  & 1005$\pm$109\\
    &N~{\sc v}$\lambda$1240                             & 10.5$\pm$0.7  & 4855.16$\pm$0.09 & 32.66$\pm$1.08 & 1925$\pm$70  & 77.64$\pm$8.34\\
    & O~{\sc i}+Si~{\sc ii}$\lambda$1305                & 0.67$\pm$0.15 & 5108.32$\pm$0.17 & 14.72$\pm$0.56 & 621$\pm$46   & 9.11$\pm$0.76\\
    & Si~{\sc iv}$\lambda$1397+O~{\sc iv}]$\lambda$1402 & 3.34$\pm$0.29 & 5488.64$\pm$0.20 & 45.38$\pm$2.36 & 2405$\pm$133 & 47.92$\pm$3.79\\
    & N~{\sc iv}]$\lambda$1486                          & 1.28$\pm$0.20 & 5811.57$\pm$0.89 & 39.45$\pm$4.82 & 1943$\pm$260 & 24.85$\pm$3.23\\
    & C~{\sc iv}$\lambda$1549                           & 10.3$\pm$0.4  & 6062.13$\pm$0.06 & 45.03$\pm$0.27 & 2145$\pm$14  & 148.8$\pm$3.5\\
    &He~{\sc ii}$\lambda$1640                           & 8.98$\pm$0.30 & 6416.40$\pm$0.09 & 28.13$\pm$0.54 & 1169$\pm$28  & 156.1$\pm$11.5\\
    &O~{\sc iii}]$\lambda$1663                          & 2.13$\pm$0.25 & 6510.19$\pm$0.34 & 57.33$\pm$3.02 & 2571$\pm$143 & 44.63$\pm$4.53\\
    &C~{\sc iii}]$\lambda$1909                          & 7.41$\pm$0.61 & 7464.27$\pm$0.23 & 43.17$\pm$3.04 & 1626$\pm$130 & 173.5$\pm$32.0\\
    &[O~{\sc iii}]+C~{\sc ii}+Si~{\sc ii}$\lambda$2322  & 0.96$\pm$0.21 & 9106.79$\pm$0.35 & 23.55$\pm$0.76 & 490$\pm$40   & 31.74$\pm$2.79\\
    \hline
    USS 1545-234
    &Ly$\alpha\lambda$1216                              & 18.0$\pm$0.2  & 4563.85$\pm$0.01 & 15.23$\pm$0.12 & 801$\pm$10  & 1315$\pm$465\\
    &N~{\sc v}$\lambda$1240                             & 7.01$\pm$0.48 & 4646.44$\pm$0.16 & 37.71$\pm$0.91 & 2358$\pm$60 & 85.61$\pm$5.74\\
    & Si~{\sc iv}$\lambda$1397+O~{\sc iv}]$\lambda$1402 & 1.30$\pm$0.14 & 5249.85$\pm$0.06 & 23.54$\pm$1.19 & 1202$\pm$76 & 39.17$\pm$6.87\\
    & C~{\sc iv}$\lambda$1549                           & 6.76$\pm$0.30 & 5810.36$\pm$0.02 & 24.77$\pm$0.31 & 1128$\pm$18 & 322.7$\pm$29.4\\
    &He~{\sc ii}$\lambda$1640                           & 4.72$\pm$0.20 & 6150.74$\pm$0.01 & 16.65$\pm$0.16 & 547$\pm$12  & 191.7$\pm$14.1\\
    &O~{\sc iii}]$\lambda$1663                          & 0.85$\pm$0.13 & 6237.76$\pm$0.16 & 21.29$\pm$1.30 & 828$\pm$78  & 38.72$\pm$5.06\\
    &C~{\sc iii}]$\lambda$1909                          & 2.88$\pm$0.21 & 7151.00$\pm$0.07 & 26.68$\pm$1.15 & 944$\pm$57  & 155.6$\pm$21.2\\
    &[O~{\sc iii}]+C~{\sc ii}+Si~{\sc ii}$\lambda$2322  & 1.13$\pm$0.31 & 8730.34$\pm$0.16 & 19.02$\pm$0.53 & 322$\pm$41  & 70.83$\pm$6.08\\
    &[Ne~{\sc iv}]$\lambda$2424                         & 1.72$\pm$0.24 & 9088.72$\pm$0.33 & 22.82$\pm$1.73 & 446$\pm$105 & 141.6$\pm$39.8\\
    \hline
  \end{tabular}
  \label{tab:line_flux}
  \tablecomments{$^a$Central wavelength. $^b$Observed wavelength width in FWHM before the correction for the instrumental broadening. $^c$Velocity width in FWHM after the correction for the instrumental broadening.}
\end{table*}

%%%%%%%%%%%%%%%%%%%%%%%%%%%%%%%%%%%%%%%%%%%%%%%%%%%%%%%%%%%%%%%%%%%%%%%%%%%%%%%%%%%%%%%%%5
\subsection{Rest-frame UV spectra of additional data}
The compiled fluxes of rest-frame UV emission lines of HzRGs described in Section 2.2 are listed in Table~\ref{tab:add-targets}.
The UV spectrum of 0828$+$193 shows the largest number of emission lines among our sample \citep[for details, see Table 4 in][]{2008MNRAS.383...11H}. By combining the emission-line flux ratios of our own sample (Section 3.1) and also the objects taken from the literature (Section 3.2), we discuss the physical and chemical properties of NLRs in HzRGs combined with photoionization models whose details are explained in the next section.

\begin{table*}[ht]
  \centering
  \caption{Data of additional targets}
  \begin{tabular}{lclc}\hline
    Name & $z$ & available emission lines & Reference$^a$ \\ \hline
    0211$-$122 & 2.340 & Si~{\sc iv} $+$ O~{\sc iv}], N~{\sc iv}], C~{\sc iv}, He~{\sc ii}, O~{\sc iii}], C~{\sc iii}], [Ne~{\sc iv}] & H08\\
    0406$-$244 & 2.440 & Si~{\sc iv} $+$ O~{\sc iv}], C~{\sc iv}, He~{\sc ii}, O~{\sc iii}], C~{\sc iii}], [Ne~{\sc iv}] & H08\\
    0731$+$438 & 2.429 & Si~{\sc iv} $+$ O~{\sc iv}], N~{\sc iv}], C~{\sc iv}, He~{\sc ii}, O~{\sc iii}], C~{\sc iii}], [Ne~{\sc iv}] & V01\\
    0828$+$193 & 2.572 & Si~{\sc iv} $+$ O~{\sc iv}], N~{\sc iv}], C~{\sc iv}, He~{\sc ii}, O~{\sc iii}], C~{\sc iii}], [Ne~{\sc iv}] & H08\\
    0943$-$242 & 2.922 & Si~{\sc iv} $+$ O~{\sc iv}], N~{\sc iv}], C~{\sc iv}, He~{\sc ii}, O~{\sc iii}], C~{\sc iii}] & V01\\              
    1558$-$003 & 2.479 & Si~{\sc iv} $+$ O~{\sc iv}], N~{\sc iv}], C~{\sc iv}, He~{\sc ii}, O~{\sc iii}], C~{\sc iii}], [Ne~{\sc iv}] & H08\\
    3C 256 & 1.824 & Si~{\sc iv} $+$ O~{\sc iv}], C~{\sc iv}, He~{\sc ii}, O~{\sc iii}], C~{\sc iii}], [Ne~{\sc iv}] & S99\\               
    4C$-$00.54 & 2.360 & Si~{\sc iv} $+$ O~{\sc iv}], C~{\sc iv}, He~{\sc ii}, O~{\sc iii}], C~{\sc iii}], [Ne~{\sc iv}] & H08\\
    4C$+$23.56 & 2.470 & Si~{\sc iv} $+$ O~{\sc iv}], N~{\sc iv}], C~{\sc iv}, He~{\sc ii}, O~{\sc iii}], C~{\sc iii}], [Ne~{\sc iv}] & H08\\
    4C$+$40.36 & 2.265 & Si~{\sc iv} $+$ O~{\sc iv}], N~{\sc iv}], C~{\sc iv}, He~{\sc ii}, O~{\sc iii}], C~{\sc iii}], [Ne~{\sc iv}] & H08\\
    4C$+$48.48 & 2.343 & Si~{\sc iv} $+$ O~{\sc iv}], N~{\sc iv}], C~{\sc iv}, He~{\sc ii}, O~{\sc iii}], C~{\sc iii}], [Ne~{\sc iv}] & V01\\
    NVSS J002402$-$325253 & 2.043 & Si~{\sc iv} $+$ O~{\sc iv}], C~{\sc iv}, He~{\sc ii}, O~{\sc iii}], C~{\sc iii}], [Ne~{\sc iv}] & DB06\\                
    \hline
  \end{tabular}
  \label{tab:add-targets}
  \tablecomments{$^a$S99 = \citet{1999ApJ...525..659S}; V01 = \citet{2001A&A...366....7V}; DB06 = \citet{2006MNRAS.366...58D}; H08 = \citet{2008MNRAS.383...11H}. $^2$Redshifts from the NASA/IPAC Extragalactic Database (NED).}
\end{table*}

%%%%%%%%%%%%%%%%%%%%%%%%%%%%%%%%%%%%%%%%%%%%%%%%%%%%%%%%%%%%%%%%%%%%%%%%%%%%%%%%%%%%%%%%%%%
\section{Photoionization model fitting and results}
\subsection{Method}
As mentioned in Section 1, the main ionization mechanism of NLR clouds in AGNs has been thought to be the photoionization by ionizing photons from the central engine of AGNs \citep[e.g.,][]{1996A&A...312..365B,1997A&A...323...31K,2004ApJS..153...75G,2016ApJ...833..266T}.
On the other hand, the collisional ionization of NLR clouds by fast shocks associated with radio jets and outflows has been also suggested for some AGNs \citep[e.g.,][]{1996AJ....112...81K,1998ApJ...495..680B,2008ApJS..178...20A,2013ApJ...772..138S,2016ApJ...833..190T}.
\citet{2009A&A...503..721M} reported that most gas clouds (especially clouds emitting high-ionization emission lines, see Table~\ref{tab:IPs}) in NLRs of HzRGs are photoionized, based on the C~{\sc iii}]/C~{\sc iv} versus C~{\sc iv}/He~{\sc ii} diagnostic diagram \citep[see also][]{2006A&A...447..863N}.
  Therefore, in this work, physical and chemical properties of ionized gas clouds in NLRs are investigated through the comparison between photoionization models and observed emission-line spectra.

\begin{table}[htb]
  \centering
  \caption{Ionization potentials of ions}
  \begin{tabular}{lc}\hline\hline
    Ion & ionization potential (eV) \\
    \hline
    C~{\sc iii}  & 24.4\\
    O~{\sc iii}  & 35.1\\
    Si~{\sc iv}  & 33.5\\
    N~{\sc iv}   & 47.4\\
    C~{\sc iv}   & 47.9\\
    He~{\sc ii}  & 54.4\\
    O~{\sc iv}   & 54.9\\
    Ne~{\sc iv}  & 63.5\\
    \hline
  \end{tabular}
  \label{tab:IPs}
\end{table}
 
We calculated the flux ratio based on photoionization model using Cloudy version 13.03 \citep{2013RMxAA..49..137F}. We used the \texttt{table AGN} command as the input SED, which reproduces the typical ionizing SED of AGNs \citep{1987ApJ...323..456M}.
The parameter ranges covered in the model calculations were the hydrogen gas density log~$n$ = 2.0 $-$ 6.0, ionization parameter log~$U = -3.0- -0.5$, and metallicity $Z =$ 0.1$-$5.0 $Z_{\odot}$. The step of calculations was 0.1 dex for each parameter, and thus 53300 models were calculated.
The adopted ranges of these parameters are typical for NLRs, which have been adopted in the literature \citep[e.g.,][]{2006A&A...447..863N,2016MNRAS.456.3354F}.
The relative elemental abundance ratio is assumed to be the solar composition \citep{2010Ap&SS.328..179G}, except for helium and nitrogen. For helium, we take the primordial component and the primary nucleosynthesis component into account. Specifically, it is determined by He/H $= 0.08096 + 0.02618$($Z/Z_{\odot}$).
The nitrogen relative abundance is assumed to be proportional to the square of the metallicity at the high-metallicity range, taking the nature of nitrogen as a secondary element into account.
  More specifically, the nitrogen relative elemental abundance is determined by log (N/H) $ = -4.57+$log$(Z/Z_{\odot})$ for 0.1 and 0.2~$Z_{\odot}$ models and log (N/H) $ = -4.17+2$log$(Z/Z_{\odot})$ for $Z \geq 0.3$ $Z_{\odot}$ models.
These analytic expressions adopted in this work for helium and nitrogen were taken from \citet{2000ApJ...542..224D}.
We assumed dust-free gas clouds in our Cloudy runs, i.e., relative elemental abundance of gas clouds is without dust depletion.
This is because high-ionization lines arise mostly in the inner part of the NLR, where dust grains do not survive \citep{1994A&A...291...18M,2003AJ....125.1729N,2006A&A...447..863N}.
The model calculations were terminated when the ionized fraction of hydrogen drops to 15~\%, because below which the gas does not emit rest-frame UV emission lines significantly. 

  Although the model runs in this work were executed by assuming one-zone constant-density clouds, it has been reported that high-ionization lines and low-ionization lines arise at different parts in the NLR with a significantly different gas density \citep[e.g.,][]{1997ApJ...487..122F,2001ApJ...549..155N,2015MNRAS.448.2900R,2016ApJ...831...68A}. In this work, we focus only on high-ionization lines (Si~{\sc iv}, O~{\sc iv}], N~{\sc iv}], C~{\sc iv}, He~{\sc ii}, O~{\sc iii}], C~{\sc iii}], and [Ne~{\sc iv}]), that are expected to arise from a similar part within NLRs (Table~\ref{tab:IPs}).
Therefore it is expected that our one-zone treatment does not introduce significant uncertainty.

%%%%%%%%%%%%%%%%%%%%%%%%%%%%%%%%%%%%%%%
\subsection{Model results}
We determine the model parameters (the gas density, ionization parameter, and gas metallicity) simultaneously, by the $\chi$ square ($\chi^{2}$) fitting where the reduced $\chi^{2}$ ($\tilde{\chi}^{2}$) is calculated. 
All line fluxes are normalized by the He~{\sc ii} flux, because He~{\sc ii} is a recombination line and thus its flux is almost proportional to the number of the He$^+$-ionizing photons without significant dependences on ISM properties such as the gas density and ionization parameter.
Note that, though the relative He$^+$-ionizing photon luminosity depends on the SED, the UV SEDs of AGNs are similar to each other if the Eddington ratio is moderately large, $\gtrsim$ 0.01 \citep[e.g.,][]{1997ApJ...475..469Z,1999PASP..111....1K,2006ApJS..166..470R}.

Here it should be noted that the error of observed emission-line fluxes given in Table~\ref{tab:line_flux} does not include the systematic error of flux measurements. Indeed, the signal-to-noise ratios of the strong emission lines are higher than that of faint emission lines.
In other words, the weight of faint emission lines is small in the $\chi^2$ fitting.
Then, the following systematic errors are considered to exist, and the weight of the strong emission lines is reduced by add those systematic errors to the error of observed emission-line fluxes.
  FORS2 Absolute Photometry Project reported that photometric accuracy of 1$\sigma$ systematic error is $\sim$ 2\% on a photometric condition\footnote{VLT-TRE-ESO-13112-5727 (https://www.eso.org/sci/facilities/\\paranal/instruments/fors/doc/VLT-TRE-ESO-13112-5727.pdf)}.
  In general, the absolute photometric accuracy of long-slit spectroscopic observations gets worse than that of photometric observations because of uncertainty of acquisition onto slit position and slit-loss depending on the seeing condition.
  Moreover our observations were mostly performed in worse seeing conditions in which typical seeing was 1$^{\prime\prime}$.4.
  Thus the measurement accuracy was worse due to slit loss.
  As another error cause, we measured the line fluxes by fitting with a single Gaussian profile to each line.
  However, some emission lines show asymmetric profiles and deviations from the ideal Gaussian profile (Figures \ref{fig:j0920}--\ref{fig:uss1545}), and thus measured fluxes are incomplete to represent the actual fluxes.
  In order to correct these errors, we conservatively adopt the 1$\sigma$ flux error to be 10\% for emission lines detected with SN $>$ 10, in the following $\chi^{2}$ calculations.

We perform the minimum $\tilde{\chi}^2$ search 10000 times using observed emission-line flux ratios. In each search, we vary the emission-line flux ratios randomly within 1$\sigma$ flux error of the observed line-flux ratios to evaluate the scatter of the ISM parameters.
The best-fit parameters are determined by the median value of the minimum $\tilde{\chi}^2$ in the search.
The errors of parameters are estimated by the 18 and 84 percentile values of the distribution of the minimum $\chi^2$.
The obtained median of the minimum $\tilde{\chi}^2$ and best-fit parameters are summarized in Table~\ref{tab:M09_results} for 15 HzRGs in our sample.

\begin{table*}[htb]
  \centering
  \caption{Comparison between photoionization models and observations}
  \begin{tabular}{lcccc}\hline\hline
    Name & $\tilde{\chi}^2$ & $Z$ ($Z_{\odot}$) & log $n$ (cm$^{-3}$)& log $U$\\
    \hline
    TN J0920$-$0712 & 3.79 & 1.2$^{+0.1}_{-0.2}$ & 4.4$^{+0.2}_{-0.2}$ & $-1.7^{+0.1}_{-0.1}$\\
    4C 24.28 & 13.19 & 2.1$^{+0.2}_{-0.9}$ & 5.3$^{+0.1}_{-0.9}$ & $> -1.7$ $^a$\\
    USS 1545$-$234 & 9.93 & 1.4$^{+0.2}_{-0.2}$ & 4.3$^{+0.2}_{-0.2}$ & $-1.4^{+0.2}_{-0.1}$\\
    \hline
    0211$-$122& 12.34 & 1.4$^{+0.1}_{-0.1}$ & 3.7$^{+0.2}_{-0.2}$ & $-1.3^{+0.1}_{-0.2}$\\ %& ($z$=2.340)
    0406$-$244& 6.01 & 2.0$^{+2.7}_{-0.6}$ & 4.4$^{+1.2}_{-0.5}$ & $> -2.0$ $^a$\\%& ($z$=2.440)
    0731$+$438& 15.29 & 0.7$^{+0.0}_{-0.1}$ & $<$ 2.4 $^a$ & $-1.5^{+0.0}_{-0.1}$\\
    0828$+$193& 12.94 & 0.7$^{+0.1}_{-0.0}$ & $<$ 2.2 $^a$ & $-1.4^{+0.1}_{-0.1}$\\%& ($z$=2.572)
    0943$-$242& 7.39 & 0.9$^{+0.0}_{-0.1}$ & $<$ 2.3 $^a$ & $-1.6^{+0.1}_{-0.1}$\\ 
    1558$-$003& 5.51 & 0.7$^{+0.2}_{-0.1}$ & 4.2$^{+0.3}_{-0.6}$ & $-1.5^{+0.1}_{-0.1}$\\%& ($z$=2.479)
    3C 256& 27.05 & 1.3$^{+0.1}_{-0.3}$ & 4.1$^{+0.2}_{-0.3}$ & $-1.8^{+0.1}_{-0.1}$\\
    4C$-$00.54& 31.32 & 1.4$^{+0.1}_{-0.2}$ & 4.2$^{+0.2}_{-0.3}$ & $-1.0^{+0.1}_{-0.2}$\\%& ($z$=2.360)
    4C$+$23.56& 18.44 & 1.1$^{+0.3}_{-0.2}$ & 3.3$^{+0.7}_{-0.4}$ & $-1.5^{+0.1}_{-0.1}$\\%& ($z$=2.470)
    4C$+$40.36& 10.99 & 0.6$^{+0.0}_{-0.1}$ & $<$ 2.1 $^a$ & $-1.9^{+0.1}_{-0.0}$\\%& ($z$=2.265)
    4C$+$48.48& 13.80 & 1.3$^{+0.2}_{-0.1}$ & 3.9$^{+0.2}_{-0.3}$ & $-1.6^{+0.1}_{-0.1}$\\
    J0024$-$3252& 5.12 & 1.6$^{+0.3}_{-0.2}$ & 4.3$^{+0.1}_{-0.2}$ & $-1.8^{+0.1}_{-0.1}$\\%& ($z$=2.043)
    \hline
  \end{tabular}
  \tablecomments{$^a$The cases reaching the upper/lower limit of parameter ranges in our model runs.}
  \label{tab:M09_results}
\end{table*}

\begin{figure*}[htb]
  \begin{center}
    \epsscale{0.9}
    \plotone{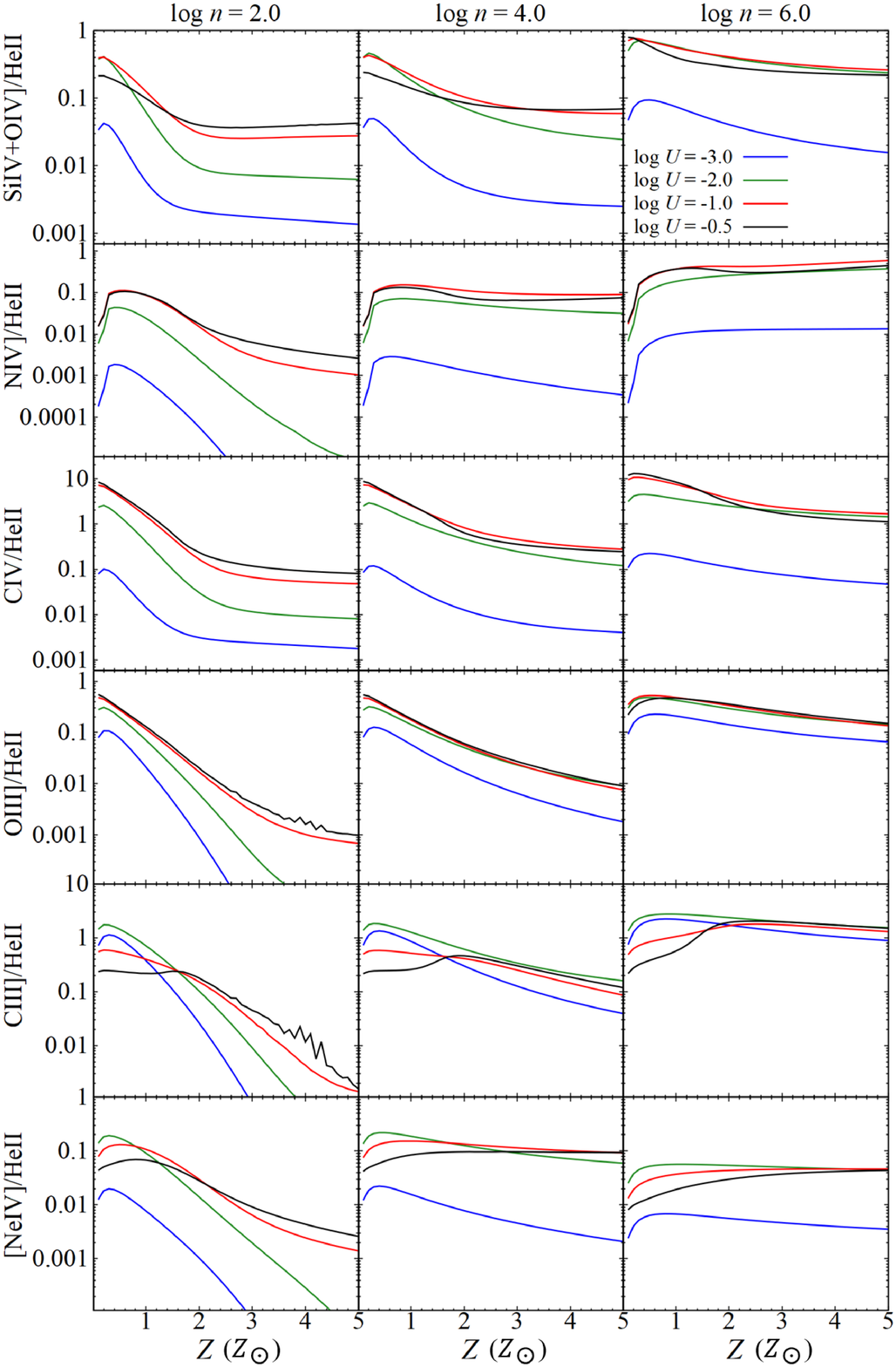}
    \caption{The relationship between the predicted flux ratio and metallicity. From left to right, models with log~$n$ = 2, 4, and 6 are shown. The blue, green, red, and black lines denote the results with log~$U = -3, -2, -1$, and $-0.5$, respectively.}
    \label{fig:model_ratio}
  \end{center}
\end{figure*}

In order to understand which emission-line ratio is the most important to determine each ISM parameter, we check the behavior of each flux ratio as a function of the ISM parameters.
Figure~\ref{fig:model_ratio} shows the dependence of the flux ratios on the metallicity, where the models are calculated with parameter sets of log~$U = -3.0, -2.0, -1.0$, and $-0.5$, and log~$n =$ 2.0, 4.0, and 6.0.
All line ratios increase with increasing the gas density from log~$n$ = 2.0 to 6.0 (i.e., from left to right panel), especially at models with a higher metallicity.
These trends suggest that the gas density is closely related to almost all line ratios.
This density dependence of C~{\sc iv}/He~{\sc ii} in NLRs has been reported in earlier works \citep[e.g.,][]{2006A&A...447..863N}.
It is natural that the [Ne~{\sc iv}]/He~{\sc ii} line ratio shows a positive dependence on the gas density, since the [Ne~{\sc iv}] line is a forbidden emission with a critical density of log~$n_{\rm cr} \sim 5.0$ (actually the [Ne~{\sc iv}] line is an unresolved doublet emission of [Ne~{\sc iv}]$\lambda$2422 and [Ne~{\sc iv}]$\lambda$2425 with log~$n_{\rm cr} =$ 5.4 and 4.9, respectively; see, e.g., \citealt{1988ApL&C..27..275Z}).
All line ratios also show the dependence on metallicity at the low gas density models (log~$n = 2.0$), while metallicity dependency seems to be weaker at the higher gas density models.
This trend infers that gas density and metallicity are degenerated; i.e., a large flux ratio is due to either of a high gas density or a high metallicity, or both.
Thus the constraint on the gas density is important to accurately estimate the metallicity.
Figure~\ref{fig:model_ratio} also suggests that the ionization parameter is constrained by combining Si~{\sc iv}$+$O~{\sc iv}]/He~{\sc ii}, N~{\sc iv}]/He~{\sc ii}, C~{\sc ii}/He~{\sc ii}, and [Ne~{\sc iv}]/He~{\sc ii}, i.e., flux ratios of two emission lines with different ionization potentials.
These line ratios increase with a decreasing ionization parameter.

The dependence of the gas density on C~{\sc iii}]/He~{\sc ii} flux ratio is thought to be small, while that of C~{\sc iv}/He~{\sc ii} is not small \citep[e.g.,][]{2006A&A...447..863N,2009A&A...503..721M}.
Although these emission-line flux ratios have been used to discuss the metallicity of NLR clouds by assuming a certain gas density, the uncertainty of gas density in estimation of the metallicity had not been discussed in detail.
Figure~\ref{fig:model_ratio} shows that the C~{\sc iv}/He~{\sc ii} flux ratio depends on the gas density at most 2 dex from log~$n$ = 2 to 6 at the highest metallicity model, and thus the previous metallicity measurement may be affected by this density dependence.
For example, the C~{\sc iv}/He~{\sc ii} flux ratio of TN J0920$-$0712 is 1.617.
This corresponds to $\sim$~1.5~$Z_{\odot}$ assuming a model with log~$n = 4.0$ and log~$U = -1$, while metallicity is inferred to be $\sim$~5~$Z_{\odot}$ with the log~$n = 6.0$ model.
In this work, the density is well determined with an accuracy within 0.3 dex in most cases thanks to constrains from many weak emission lines, and thus the metallicity of each target can be estimated with a smaller uncertainty than previous studies based on a few emission lines. For example, \cite{2009A&A...503..721M} estimated the NLR metallicity with 0.2--0.8 dex uncertainty, even in averaged values.

%%%%%%%%%%%%%%%%%%%%%%%%%%%%%%%%%%%%%%%%
\section{Discussion}
\subsection{Interpretations of model fitting}
The best-fit photoionization models suggest that many objects in our sample show higher gas metallicities than the solar metallicity ($Z \gtrsim 1.0$ $Z_{\odot}$). These inferred metallicities are consistent with earlier works for HzRGs \citep{2000A&A...362..519D,2006A&A...447..863N,2009A&A...503..721M}.
This result suggests that the NLRs of HzRGs in our sample are already chemically matured even at $z > 3$.
The inferred ionization parameter is $-2.0 <$ log~$U < -1.0$ in our sample. \citet{2000A&A...362..519D} and \citet{2009MNRAS.395.1099B} reported that the ionization parameter of HzRGs is in the range of $-2.5 <$ log~$U < -1.5$ using rest-UV emission line ratios and photoionization models assuming the solar metallicity and gas density (log~$n = 2.0$ and 3.0).
The ionization parameter in our work distributes in the range slightly higher than the range reported in those previous works, but the difference is insignificant within the uncertainty of the estimation.
The inferred gas density of HzRGs is widely distributed in the range of log~$n = 2.0-5.0$ with a relatively large uncertainty than the derived uncertainty in the ionization parameter and gas metallicity.
The inferred range of the gas density is consistent with the range of the gas density assumed in the previous works \citep[e.g.,][]{2000A&A...362..519D,2006A&A...447..863N,2009A&A...503..721M}.
\cite{2008MNRAS.383...11H} determined the electron density of 0731$+$438, log~$n_{\rm e} < 3.5$, by using [Si~{\sc iii}]$\lambda$1883/Si~{\sc iii}]$\lambda$1892.
This result is consistent with our results (log~$n_{\rm e} < 2.4$).
    
   \cite{2018MNRAS.474.3649S} examined photoionization models for studying the rest-frame UV and optical emission-line spectrum of 0943$-$242, which is included in our sample.
   Their best-fit parameters are $Z = 2.1$ $Z_{\odot}$ and log~$U = -1.74$.
  The derived ionization parameter is close to the best-fit value in our analysis (log~$U = -1.6^{+0.1}_{-0.1}$) while the derived metallicity is higher than our measurement ($Z = 0.9^{+0.0}_{-0.1}$ $Z_{\odot}$).
  Note that, in their one-zone cloud models, the hydrogen gas density was fixed to log~$n = 2.0$.
  However Figure~\ref{fig:model_ratio} shows that most flux ratios of rest-frame UV emission lines show significant dependence on the gas density, and thus the fixed gas density may result in a possibly large systematic error in the estimates of parameters such as the metallicity.
  In addition, in their fit to derive the best-fit parameters, they used not only high-ionization lines but also some low-ionization lines (C~{\sc ii}]$\lambda$2326, Mg~{\sc ii}$\lambda$2798, and [O~{\sc ii}]$\lambda$3727).
    Since high-ionization lines and low-ionization lines generally arise at different parts in the NLR \citep[e.g.,][]{1997ApJ...487..122F}, comparisons of such diverse emission lines with one-zone photoionization models may introduce non-negligible systematic errors.
    These two aspects may be the reasons of the discrepancy in the metallicity estimate for 0943$-$242 between \citet{2018MNRAS.474.3649S} and our work.

As shown in Table~\ref{tab:M09_results}, the best-fit $\tilde{\chi}^2$ value in some cases is relatively large ($\gtrsim$ 20).
This large $\tilde{\chi}^2$ is caused probably due to over-simplification of our photoionization models.
In our model calculations, we assumed the constant chemical composition without dust grains, the typical AGN SED, and one-zone constant-density clouds.
Though our models focus only on relatively high-ionization lines, the models still suffer from such non-uniformity of NLRs that could makes the minimum $\tilde{\chi}^2$ values considerably high.

\begin{figure*}[tbh]
  \begin{center}
    \epsscale{1.1}
    \plotone{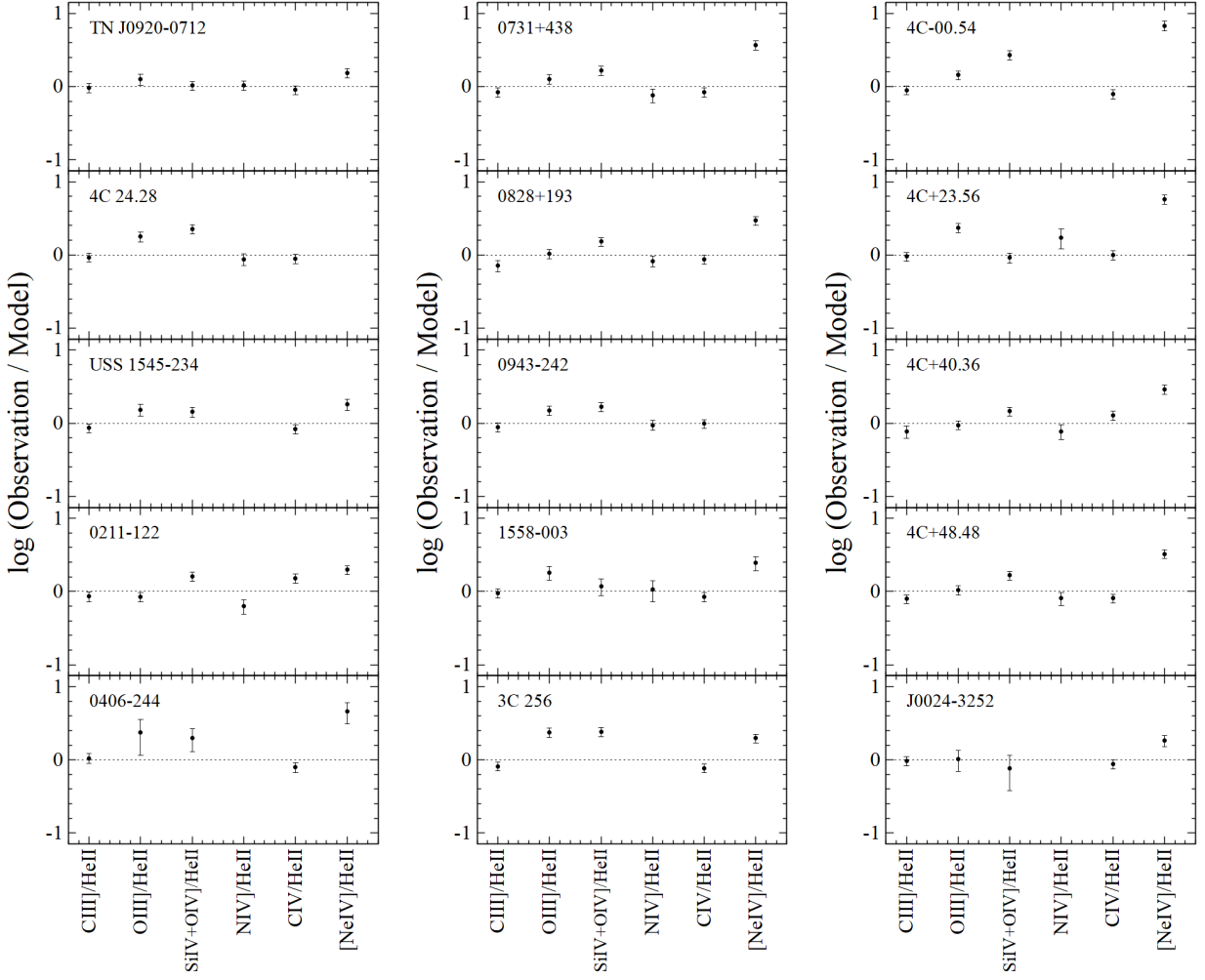}
    \caption{Comparison of each emission-line flux ratio between photoionization models and observations at the minimum $\tilde{\chi}^2$. In x-axis, the flux ratios are shown in order of increasing the ionization potential of the corresponding ion (of the numerator), from left to right. The dashed line indicates the unity, i.e., the observed flux ratios are reproduced by best-fit models.}
    \label{fig:obs_model_ratio_1}
  \end{center}
\end{figure*}

Figure~\ref{fig:obs_model_ratio_1} shows the difference in emission-line flux ratios between the observed data and the best-fit photoionization model.
In most cases, the observed C~{\sc iii}]/He~{\sc ii} and C~{\sc iv}/He~{\sc ii} flux ratios are consistent with the model predictions within the uncertainty, while [Ne~{\sc iv}]/He~{\sc ii} flux ratios are systematically under-predicted by the best-fit models.
One possible explanation of this discrepancy is the higher ionization potential of Ne~{\sc iv} (63.5 eV) compared to the other emission lines (Table~\ref{tab:IPs})\footnote{The ionization potentials shown in Table~\ref{tab:IPs} are the value required to create the ions needed to emit each emission line. Note that He~{\sc ii} is a recombination line, so the creation of the He$^{2+}$ ion is required to radiate the He~{\sc ii} line. On the other hand, the other emission lines are collisionally-excited lines, which do not require the presence of ions that are ionized to the higher ionization level.}.
Given the stratified structure of actual NLRs, there could be an additional highly-ionized clouds in the inner part of NLRs \citep[e.g.,][]{1997ApJ...487..122F,2001ApJ...549..155N,2001PASJ...53..629N,2015MNRAS.448.2900R,2016ApJ...831...68A} that are not taken into account in our one-zone photoionization model.

As mentioned in Section 2.1, we assume that the internal dust extinction is negligible for our HzRGs samples.
In order to see the effect of this assumption, we perform the $\chi^2$ fitting by using extinction-corrected emission-line flux ratios.
In this test, we assume $A_V = 0.5$, the \cite{1989ApJ...345..245C} extinction curve, and $R_V = 3.1$.
As a result, obtained best-fit parameters of the ISM are almost consistent within 1$\sigma$ error of the results without extinction correction.
This suggests that the ISM parameters inferred by the comparison with photoionization models are not sensitive to the assumption on the dust extinction.

It has been reported that the ionizing radiation of some AGNs show significant variability in a relatively short timescale ($\lesssim 10^4$ years), and consequently the ionization of NLRs in such AGNs can be largely affected (e.g., \citealt[][]{2019ApJ...870...65I}; see also \citealt[][]{2013ApJ...778...58B,2019ApJ...886...45B,2017ApJS..228...11G}).
  Such a short flare-like variability will cause inhomogeneous ionization structures of NLRs, which is hard to be described by one-zone photoionization models as given in this paper.
  Multi-zone photoionization models for NLRs \citep[e.g.,][]{1997ApJ...487..122F} may be useful to investigate such inhomogeneous structures of NLRs in high-$z$ AGNs including HzRGs.
  Since multi-zone photoionization models involve more free parameters than one-zone photoionization models generally, studies of the inhomogeneity of high-$z$ NLRs require much more emission lines (more than the lines in this paper; i.e., $>$10).
  Thus a wider spectroscopic coverage including near-infrared (i.e., rest-frame optical) will be important for detailed understandings of the ISM in high-$z$ AGNs.
%%%%%%%%%%%%%%%%%%%%%%%%%%%%%%%%%%%%%%%%%%%%%%%%%%%%%%%%%
\subsection{The nature of HzRGs}
\begin{figure}[tbh]
  \epsscale{1.2}
  \plotone{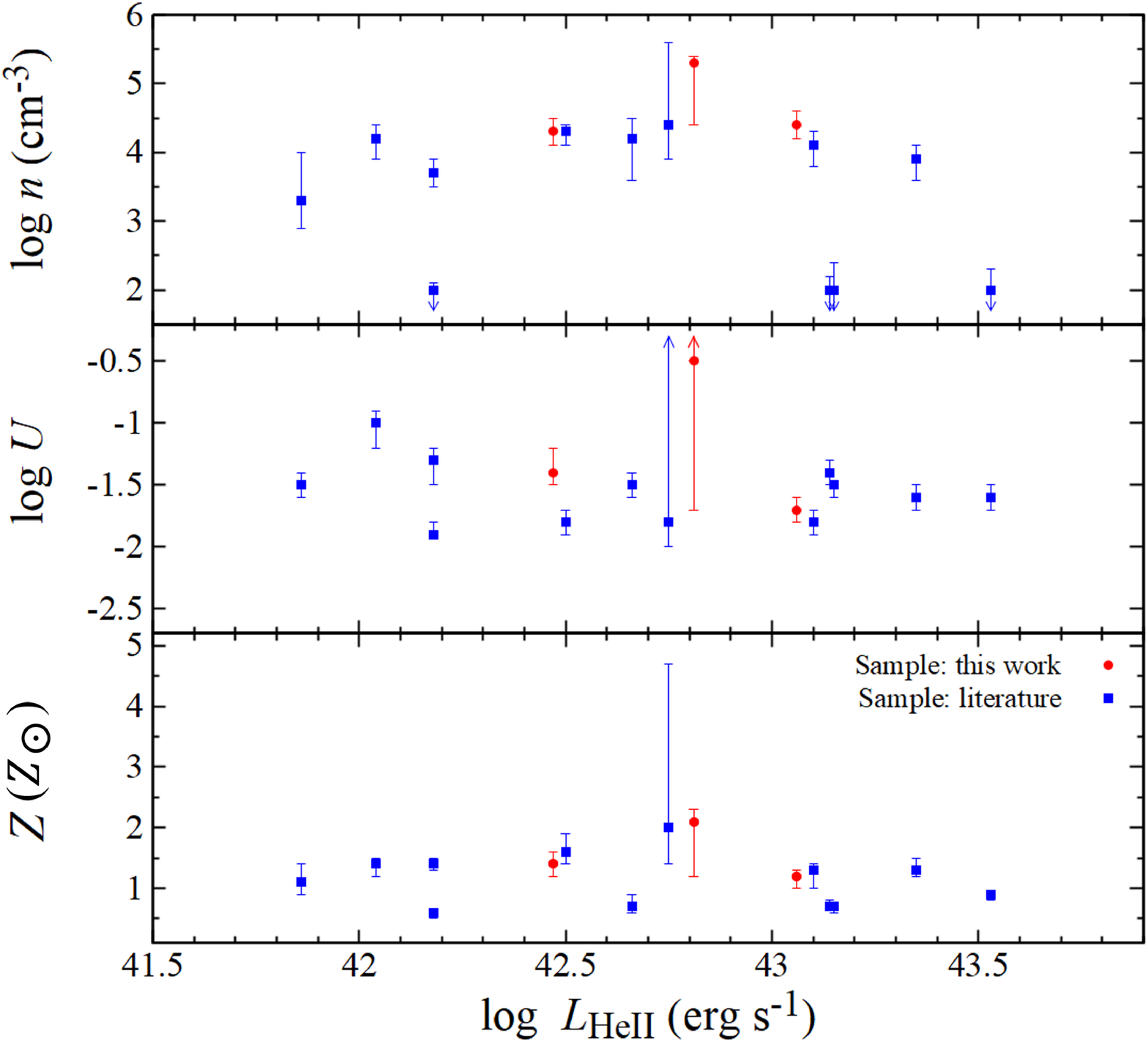}
  \caption{Inferred parameters (gas density, ionization parameter, and gas metallicity, in the upper, middle, and lower panels, respectively) as a function of the He~{\sc ii} luminosity. Red circles are our targets, while blue squares are HzRGs taken from the literature. Arrows denote the cases where the inferred error reaches the limit of calculations.}
  \label{fig:L_HeII_para}
\end{figure}

Here we investigate possible dependencies of the derived parameters of NLR clouds on the AGN luminosity.
In Table~\ref{tab:L_line}, the C~{\sc iv} and He~{\sc ii} emission-line luminosities of our sample are summarized.
Both of the two line luminosities have been often used as indicators of the AGN luminosity, but the He~{\sc ii} luminosity is a better indicator for the AGN activity (see Section 4).
The derived parameters of NLR clouds, i.e., the gas density, ionization parameter, and metallicity, are shown as functions of the He~{\sc ii} luminosity in Figure~\ref{fig:L_HeII_para}.
%%%
%%%%
From Spearman's rank-correlation test, the obtained correlation coefficients of the gas density, ionization parameter, and metallicity to the He~{\sc ii} luminosity are $-$0.18, $-$0.18, and $-$0.20, respectively.
  These values suggest that these parameters do not significantly depend on the He~{\sc ii} luminosity.
This is interesting because some previous studies reported the dependence of the NLR metallicity on the AGN luminosity \citep[e.g.,][]{2006A&A...447..863N,2009A&A...503..721M}.
Note that such luminosity dependence of the AGN metallicity has been reported also for BLR clouds \citep[e.g.,][]{1993ApJ...418...11H,2006A&A...447..157N,2011A&A...527A.100M}.
The lack of the significant luminosity dependence of the NLR metallicity in our sample is probably due to the small coverage of luminosity and small number statistics of our work; in previous works the luminosity dependence of the NLR metallicity is investigated based on averaged values of a larger sample that consists of $\sim$50 HzRGs \citep[see][]{2006A&A...447..863N,2009A&A...503..721M}. The luminosity coverage of sample in \citet{2009A&A...503..721M} is 41.5 $<$ log $L_{\rm HeII} <$ 45.0 while our work misses luminous samples (43.5 $<$ log $L_{\rm HeII} <$ 45.0) that have higher metallicity than less luminous samples. As a result, the luminosity dependence may be disappeared in our sample.

\begin{figure}[tbh]
  \centering
  \epsscale{1.22}
  \plotone{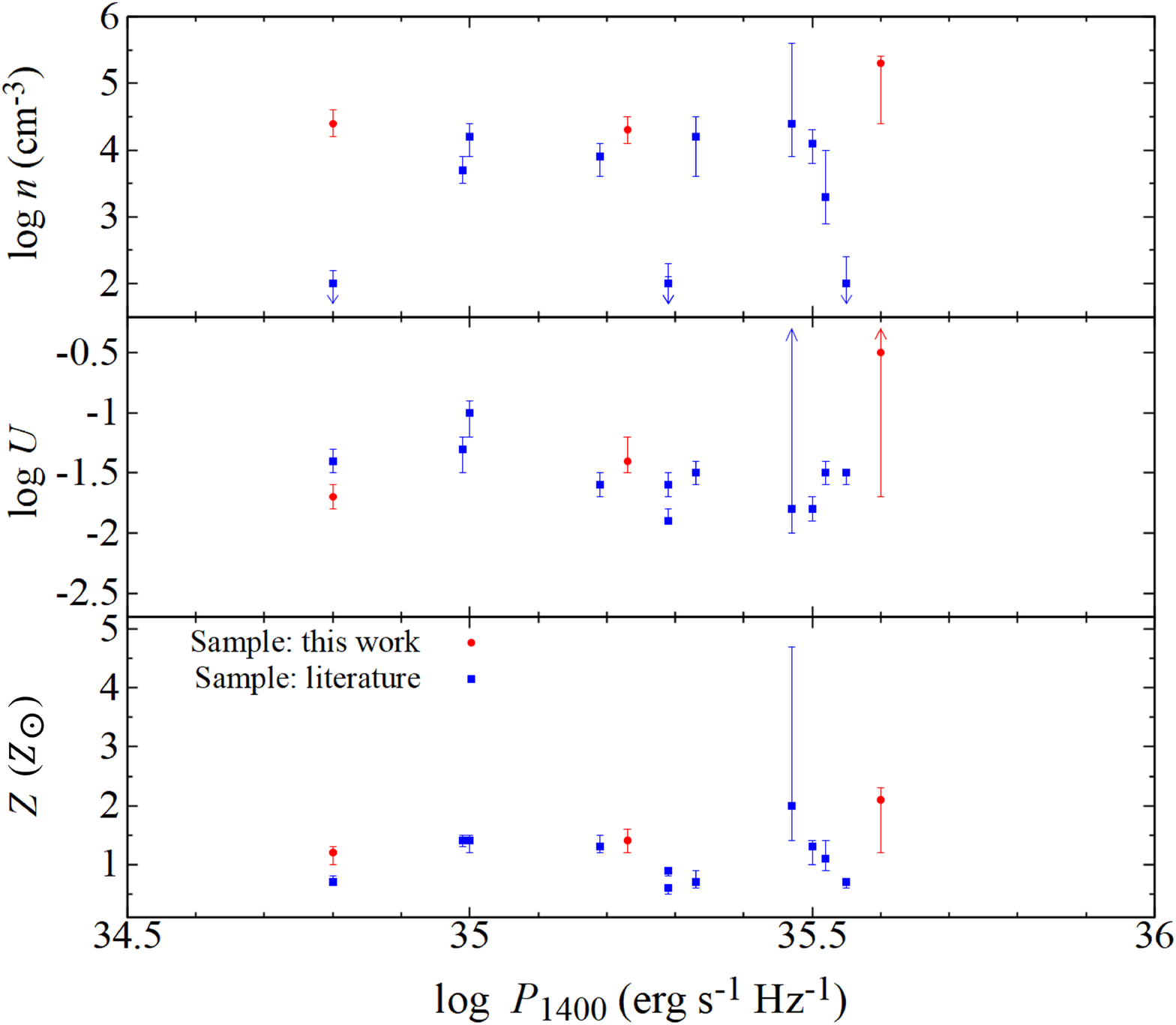}
  \caption{Relation between the radio power at 1400 MHz ($P_{1400}$) and the derived NLR parameters, i.e., the gas density (upper), ionization parameter (middle), and metallicity (lower). Red circles denote our targets. Blue squares denote additional HzRG samples from the literature. Arrows denote the cases where the inferred error reaches the limit of calculations.}
  \label{fig:L_radio_1400_para}
\end{figure}

\begin{figure}[tbh]
  \centering
  \epsscale{1.22}
  \plotone{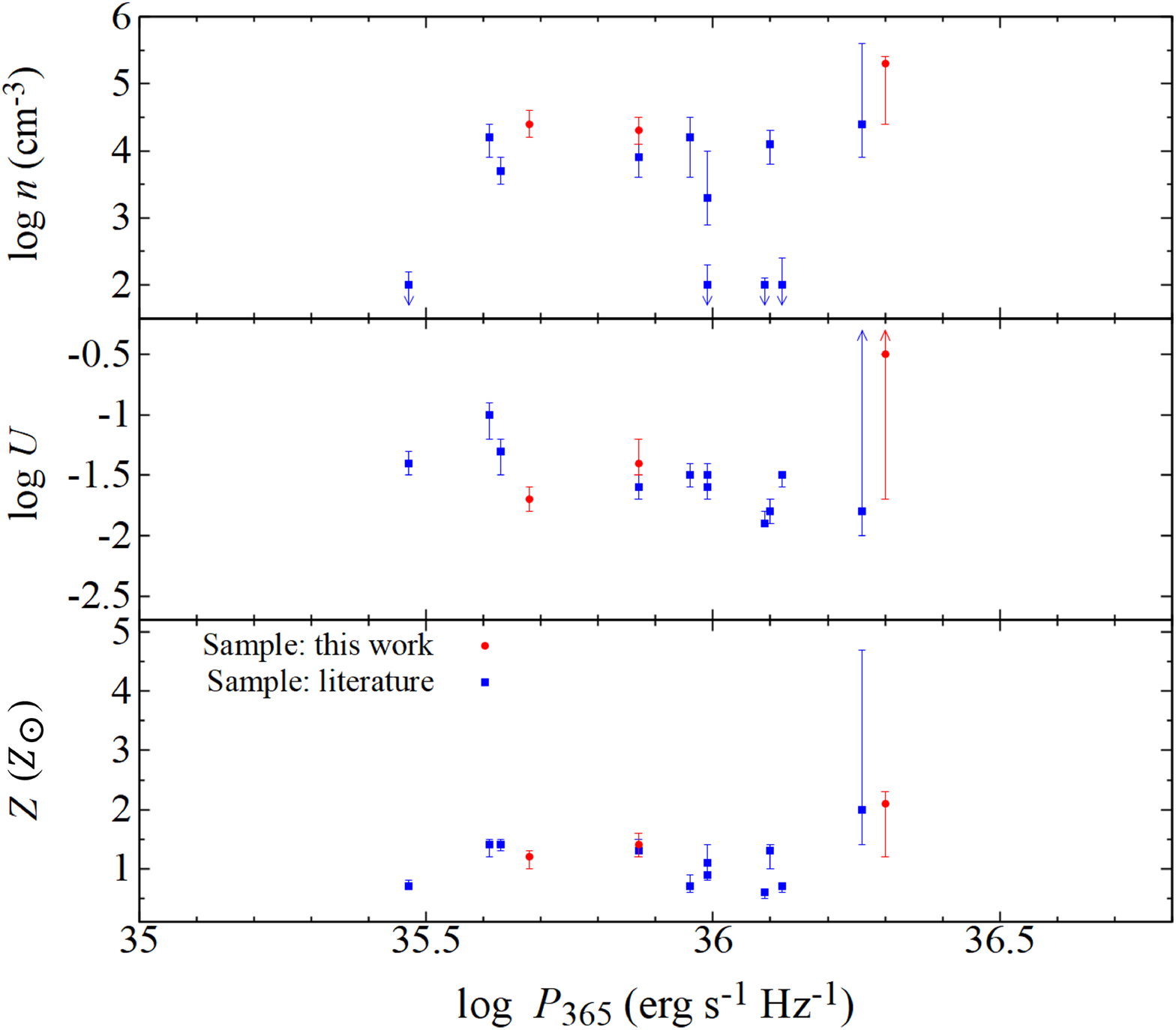}
  \caption{Same as Figure~\ref{fig:L_radio_1400_para} but for the radio power at 365 MHz ($P_{365}$).}
  \label{fig:L_radio_365_para}
\end{figure}

We also examine the dependence of the NLR parameters on the radio power, in order to investigate whether some NLR parameters are strongly affected by the radio jet in HzRGs.
For this purpose, we compiled the radio power at 365 MHz ($P_{365}$) and 1400 MHz ($P_{1400}$) summarized in Table~\ref{tab:L_line}, which are taken from \citet{1992ApJS...79..331W}, \citet{1996AJ....111.1945D}, and \citet{1998AJ....115.1693C}.
The relation between the NLR parameters and the radio power is shown in Figures~\ref{fig:L_radio_1400_para} and \ref{fig:L_radio_365_para}.
From Spearman's rank-correlation test, the obtained correlation coefficients of the gas density, ionization parameter, and metallicity to the radio power are 0.17, $-$0.37, and 0.11 for $P_{365}$, and 0.01, $-$0.22, and $-$0.05 for $P_{1400}$, respectively.
From these results, the ionization parameter might be weakly correlated with $P_{365}$.
The other ISM parameters do not show significant dependence on the radio power.
These figures and rank-correlation test suggest that there are no clear dependence of the NLR parameters on the radio power, suggesting that the physical properties of NLRs in our sample are not significantly affected by the radio jet.
This is consistent with the idea that the NLR clouds of AGNs in our samples are mostly ionized by the photoionization, not by the fast shock associated with the radio jet.
However, in order to conclude the influence of radio jets, it is necessary to investigate the spatial distributions between radio jets and line emitting clouds \citep{2000MNRAS.314..849T,2017A&A...599A.123N}.

%%%%%%%%%%%%%%%%%%%%%%%%%%%%%%%%%%%%
\begin{table*}[thb]
  \centering
  \caption{Line luminosities and radio power}
  \begin{tabular}{lccccc}\hline\hline
    Name & log $L_{\rm CIV}$  & log $L_{\rm HeII}$ & & log $P_{365}$ $^a$ & log $P_{1400}$ $^b$ \\
     & (erg s$^{-1}$) & (erg s$^{-1}$) && (erg s$^{-1}$ Hz$^{-1}$) & (erg s$^{-1}$ Hz$^{-1}$) \\ \hline
    J0920$-$0712 & 43.27$\pm$0.01 & 43.06$\pm$0.02 && 35.68$\pm$0.02 & 34.80$\pm$0.01 \\
    4C 24.28     & 42.87$\pm$0.02 & 42.81$\pm$0.01 && 36.30$\pm$0.01 & 35.60$\pm$0.01\\
    USS 1545$-$234 & 42.63$\pm$0.02 & 42.47$\pm$0.02 && 35.87$\pm$0.01 & 35.23$\pm$0.01\\
    \hline
    0211$-$122 & 42.44$\pm$0.01 & 42.18$\pm$0.01 && 35.63$\pm$0.01 & 34.99$\pm$0.02 \\
    0406$-$244 & 42.63$\pm$0.05 & 42.75$\pm$0.04 && 36.26$\pm$0.01 & 35.47$\pm$0.01 \\
    0731$+$438 & 43.33$\pm$0.01 & 43.15$\pm$0.01 && 36.12$\pm$0.01 & 35.55$\pm$0.01 \\
    0828$+$193 & 43.38$\pm$0.01 & 43.14$\pm$0.01 && 35.47$\pm$0.02 & 34.80$\pm$0.01 \\
    0943$-$242 & 43.53$\pm$0.01 & 43.53$\pm$0.01 && 35.99$\pm$0.02 & 35.29$\pm$0.01 \\
    1558$-$003 & 43.09$\pm$0.02 & 42.66$\pm$0.01 && 35.96$\pm$0.01 & 35.33$\pm$0.01 \\
    3C 256     & 43.08$\pm$0.01 & 43.10$\pm$0.01 && 36.10$\pm$0.01 & 35.50$\pm$0.01 \\
    4C$-$00.54 & 42.18$\pm$0.03 & 42.04$\pm$0.01 && 35.61$\pm$0.03 & 35.00$\pm$0.02 \\
    4C$+$23.56 & 41.94$\pm$0.05 & 41.86$\pm$0.03 && 35.99$\pm$0.02 & 35.52$^c$ \\
    4C$+$40.36 & 42.43$\pm$0.02 & 42.18$\pm$0.03 && 36.09$\pm$0.01 & 35.29$\pm$0.01 \\
    4C$+$48.48 & 43.37$\pm$0.01 & 43.35$\pm$0.03 && 35.87$\pm$0.02 & 35.19$\pm$0.01 \\
    J0024$-$3252& 42.50$\pm$0.05 & 42.50$\pm$0.05 && -- & 34.10$\pm$0.01 \\
    \hline
  \end{tabular}
  \label{tab:L_line}
  \tablecomments{$^a$ The 365 MHz flux data were collected from \citet{1996AJ....111.1945D}. \\$^b$ The 1400 MHz flux data were collected from \cite{1998AJ....115.1693C}. \\$^c$ This 1400 MHz flux density is collected from \cite{1992ApJS...79..331W}.}
\end{table*}
%%%%%%%%%%%%%%%%%%%%%%%%%%%%%%%%%%%%%%%%%%%%%%%%%%

The obtained NLR metallicity for our sample is distributed in the range of 0.5--2.1 $Z_{\odot}$ (Figure~\ref{fig:L_HeII_para}); i.e., the NLR of most HzRGs in our sample is characterized by the solar or super-solar metallicity.
This suggests that host galaxies of HzRGs are chemically matured even at $z \sim 3$, where the average and standard deviation of stellar mass of our HzRG sample (0406$-$244, 0943$-$242, 1558$-$003, 4C+23.56, and 4C+40.36) are (2.7$\pm$1.3) $\times 10^{11} M_{\odot}$ \citep{2007ApJS..171..353S,2010ApJ...725...36D}.
As a comparison of the same stellar mass of HzRGs, local galaxies with $\sim 10^{11} M_{\odot}$ have also the solar or super-solar metallicity \citep[e.g.,][]{2004ApJ...613..898T,2019A&ARv..27....3M,2020MNRAS.491..944C}.
Some earlier works \citep[e.g.,][]{2006A&A...447..863N,2009A&A...503..721M} pointed out that NLR metallicity of HzRGs is solar or supersolar, but with some assumptions on the NLR gas parameters (such as a fixed gas density).
In this work, the chemical maturity of HzRGs is confirmed with less assumptions thanks to deep spectra with weak emission lines. 
\\

%%%%%%%%%%%%%%%%%%%%%%%%%%%%%%%%%%%%%%%%%%%%%%%%%%
\subsection{Comparison with diagnostics using only strong lines}
\begin{figure*}[tbh]
  \begin{minipage}{0.33\hsize}
    \begin{center}
      \includegraphics[width=60mm]{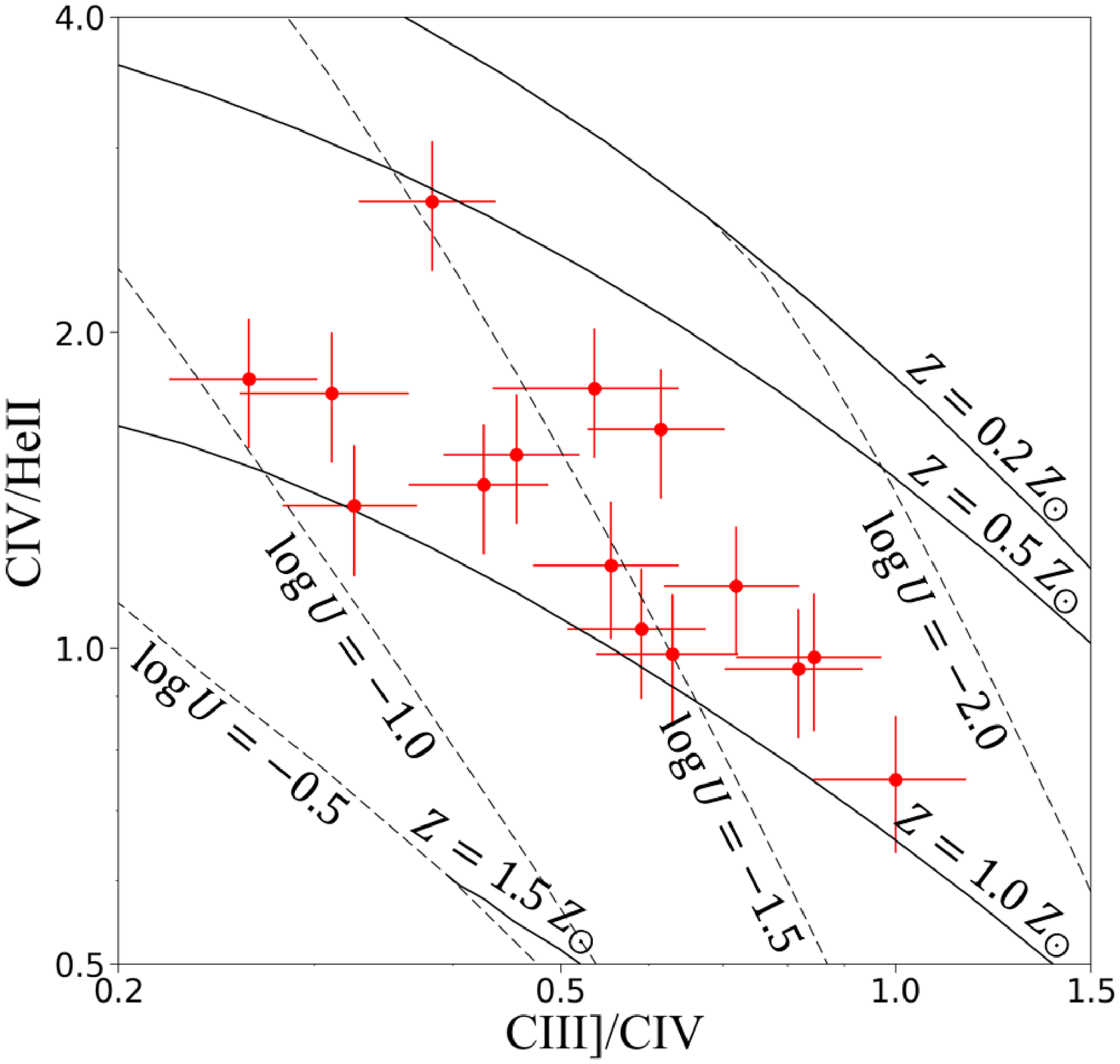}
    \end{center}
    \label{}
  \end{minipage}
  \begin{minipage}{0.33\hsize}
    \begin{center}
      \includegraphics[width=60mm]{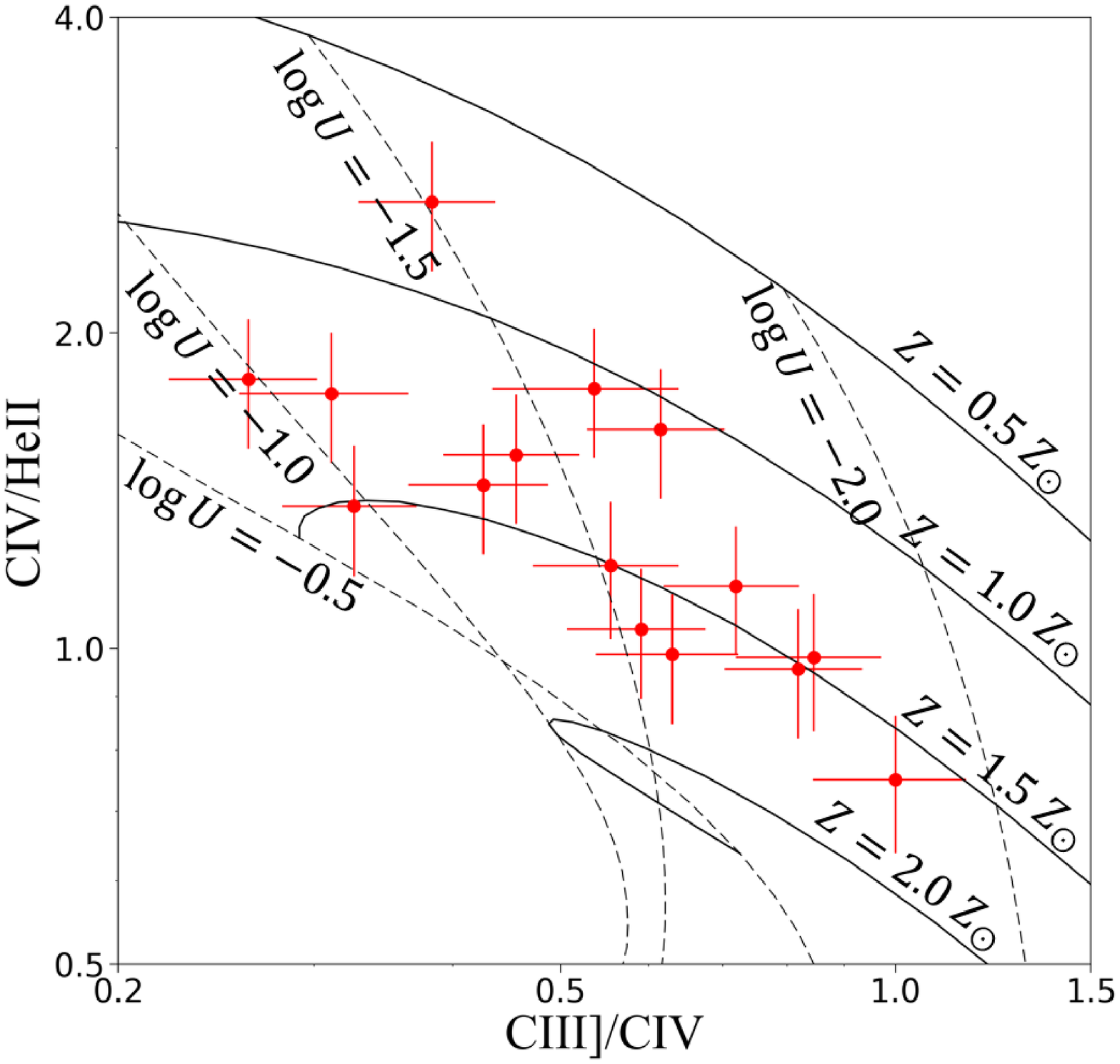}
    \end{center}
    \label{}
  \end{minipage}
  \begin{minipage}{0.33\hsize}
    \begin{center}
      \includegraphics[width=60mm]{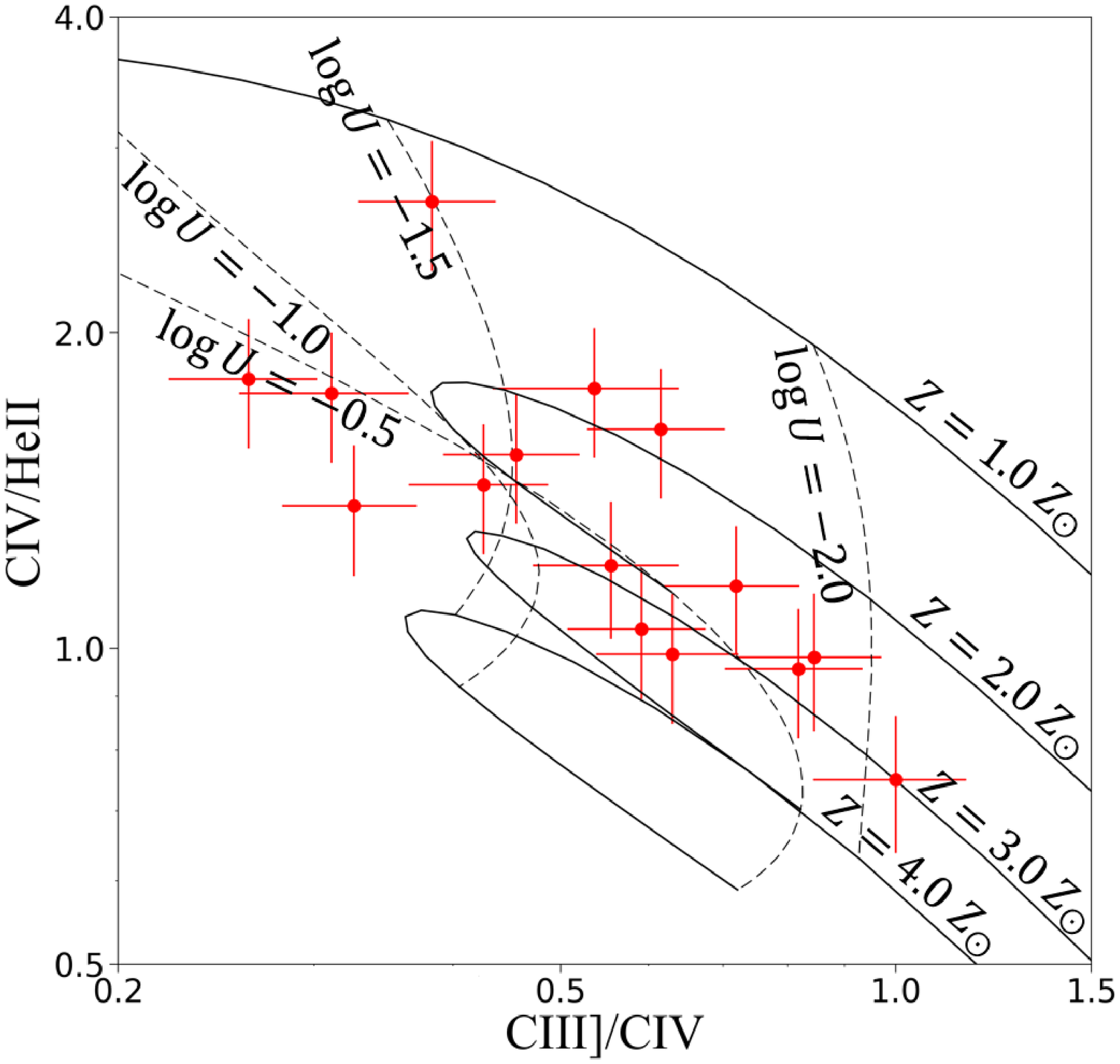}
    \end{center}
    \label{}
  \end{minipage}
  \caption{C~{\sc iii}]/C~{\sc iv} versus C~{\sc iv}/He~{\sc ii} diagram with photoionization model grids. Left, middle, and right panels show models with log $n_{\rm H} =$ 2.0, 4.0, and 5.0, respectively. Observed line ratios of our targets are plotted by red circles.}
  \label{fig:c3c4_c4he2}
\end{figure*}

\begin{figure}[tbh]
  \centering
  \epsscale{1.15}
  \plotone{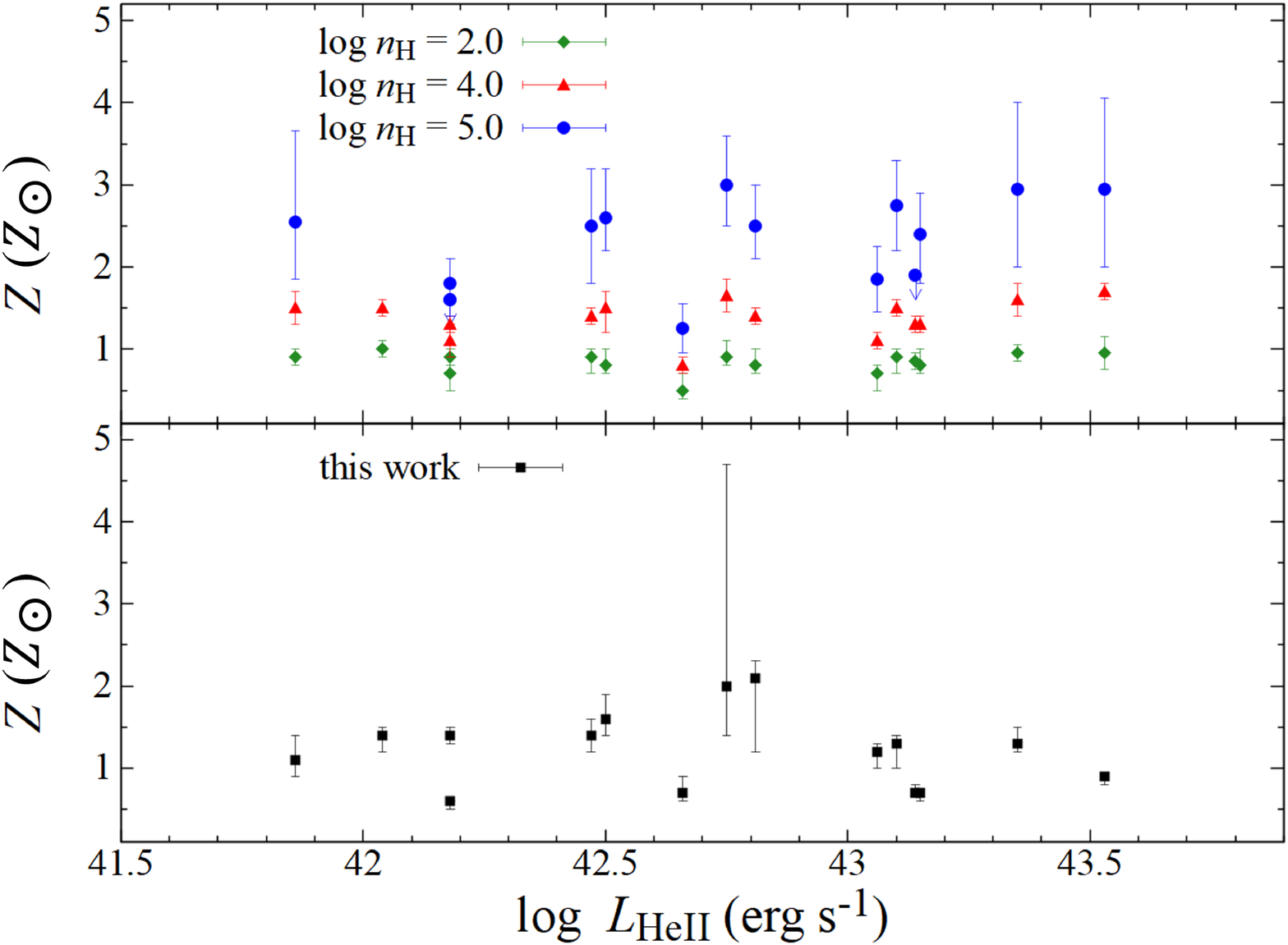}
  \caption{Inferred metallicity of our targets estimated by the line diagnostics with only strong emission lines (top panel) and by our method (bottom panel), as a function of the He~{\sc ii} line luminosity. In the top panel, green diamond, red triangle, and blue circle represent the metallicity estimated by models with log $n_{\rm H} =$ 2.0, 4.0, and 5.0, respectively.}
  \label{fig:comparison_method}
\end{figure}

To examine whether our multi-line assessment including faint lines is better than previous strong-line diagnostics, we compare the inferred NLR metallicity of HzRGs in this work (Table~\ref{tab:M09_results}) with the metallicity estimated with only strong lines.
\citet{2006A&A...447..863N} and \citet{2009A&A...503..721M} estimated the NLR metallicity by using the emission-line flux ratios of C~{\sc iv}/He~{\sc ii} and C~{\sc iii}]/C~{\sc iv}.
Figure~\ref{fig:c3c4_c4he2} shows the strong-line metallicity diagnostic diagram that consists of the flux ratios of C~{\sc iv}/He~{\sc ii} and C~{\sc iii}]/C~{\sc iv} for the cases of log $n_{\rm H}$ = 2.0, 4.0, and 5.0, with the grid of Cloudy photoionization models\footnote{Details of the model calculations are given in Section~4.1; note that we use Cloudy version 13.03 though \citet{2009A&A...503..721M} used version 07.02.}.
  The resulting metallicity based on the strong-line diagnostic diagram is given in Figure~\ref{fig:comparison_method} and Table~\ref{tab:comparison_Z}, with the metallicity estimated by our multi-line assessment (Section~4).
  Figure~\ref{fig:comparison_method} shows that the strong-line method has a large (a factor of $\sim$2--3) systematic uncertainty associated with the gas density, given the fact that we have to assume one specific density to utilize the strong-line metallicity diagnostic diagram.
  On the other hand, our multi-line assessment estimates the metallicity, gas density, and ionization parameter simultaneously.
  The resulting error is typically 0.3 dex, which is smaller than the systematic uncertainty in the strong-line method.
  Therefore we conclude that our multi-line assessment utilizing rest-UV faint emission lines is more powerful to study the NLR metallicity than previous strong-line method.
%%%%%%%%%%%%%
%%%%%%%%%%%%%
%%%%%%%%%%%%%%%
\begin{table*}[thb]
  \centering
  \caption{Estimated gas-phase matallicity$^a$ with different models}
  \begin{tabular}{lccccc}\hline\hline
    Name & This work & M+09$^b$ & M+09 & M+09 \\
     & best fit  & log $n_{\rm H}= $ 2.0  &  log $n_{\rm H}= $ 4.0  & log $n_{\rm H}= $ 5.0  \\ \hline
    J0920$-$0712 & 1.2$^{+0.1}_{-0.2}$ & 0.7$^{+0.1}_{-0.2}$ & 1.1$^{+0.1}_{-0.1}$ & 1.85$^{+0.4}_{-0.4}$ \\
    4C 24.28     & 2.1$^{+0.2}_{-0.9}$ & 0.8$^{+0.2}_{-0.1}$ & 1.4$^{+0.1}_{-0.1}$ & 2.5$^{+0.5}_{-0.4}$ \\
    USS 1545$-$234 & 1.4$^{+0.2}_{-0.2}$ & 0.9$^{+0.1}_{-0.2}$ & 1.4$^{+0.1}_{-0.1}$ & 2.5$^{+0.7}_{-0.3}$ \\
    \hline
    0211$-$122 & 1.4$^{+0.1}_{-0.1}$ & 0.9$^{+0.1}_{-0.2}$ & 1.3$^{+0.1}_{-0.1}$ & $<$ 1.6 \\
    0406$-$244 & 2.0$^{+2.7}_{-0.6}$ & 0.9$^{+0.2}_{-0.1}$ & 1.65$^{+0.2}_{-0.2}$ & 3.0$^{+0.6}_{-0.5}$ \\
    0731$+$438 & 0.7$^{+0.0}_{-0.1}$ & 0.8$^{+0.2}_{-0.1}$ & 1.3$^{+0.1}_{-0.1}$ & 2.4$^{+0.5}_{-0.6}$ \\
    0828$+$193 & 0.7$^{+0.1}_{-0.0}$ & 0.85$^{+0.1}_{-0.1}$ & 1.3$^{+0.1}_{-0.1}$ & $<$ 1.9 \\
    0943$-$242 & 0.9$^{+0.0}_{-0.1}$ & 0.95$^{+0.2}_{-0.2}$ & 1.7$^{+0.1}_{-0.1}$ & 2.95$^{+1.1}_{-0.95}$ \\
    1558$-$003 & 0.7$^{+0.2}_{-0.1}$ & 0.5$^{+0.2}_{-0.1}$ & 0.8$^{+0.1}_{-0.1}$ & 1.25$^{+0.3}_{-0.3}$ \\
    3C 256     & 1.3$^{+0.1}_{-0.3}$ & 0.9$^{+0.1}_{-0.2}$ & 1.5$^{+0.1}_{-0.1}$ & 2.75$^{+0.55}_{-0.55}$ \\
    4C$-$00.54 & 1.4$^{+0.1}_{-0.2}$ & 1.0$^{+0.1}_{-0.1}$ & 1.5$^{+0.1}_{-0.1}$ & --- \\
    4C$+$23.56 & 1.1$^{+0.3}_{-0.2}$ & 0.9$^{+0.1}_{-0.1}$ & 1.5$^{+0.2}_{-0.2}$ & 2.55$^{+1.1}_{-0.7}$ \\
    4C$+$40.36 & 0.6$^{+0.0}_{-0.1}$ & 0.7$^{+0.1}_{-0.2}$ & 1.1$^{+0.2}_{-0.2}$ & 1.8$^{+0.3}_{-0.4}$ \\
    4C$+$48.48 & 1.3$^{+0.2}_{-0.1}$ & 0.95$^{+0.1}_{-0.1}$ & 1.6$^{+0.2}_{-0.2}$ & 2.95$^{+1.05}_{-0.95}$ \\
    J0024$-$3252& 1.6$^{+0.3}_{-0.2}$ & 0.8$^{+0.2}_{-0.1}$ & 1.5$^{+0.2}_{-0.3}$ & 2.6$^{+0.6}_{-0.4}$ \\
    \hline
  \end{tabular}
  \label{tab:comparison_Z}
  \tablecomments{$^a$ Given in units of the solar metallicity.
    $^b$ \citet{2009A&A...503..721M}.}
\end{table*}
%%%%%%%%%%%%%%%%%%%%%%%%%%%%%%%%%%%%%%%%%%%%%%
\\
\section{CONCLUSION}
In this work, we focus on rest-frame UV emission lines including faint ones (such as N~{\sc iv}]$\lambda$1486, O~{\sc iii}]$\lambda$1663, and [Ne~{\sc iv}]$\lambda$2424) of 15 HzRGs at $z \sim 3$ in order to examine the ISM properties (gas density, ionization parameter, and metallicity) of NLRs in HzRGs.
We diagnose the physical and chemical properties of the ISM in NLR for each object through the comparison between the observed emission-line fluxes and detailed photoionization models. Main results of this work are as follows.
\begin{enumerate}  
\item[1.] Most HzRGs show high gas metallicity, that is close to or higher than the solar metallicities (i.e., $Z \gtrsim Z_{\odot}$). This result is consistent with some previous studies \citep[e.g.,][]{2006A&A...447..863N,2009A&A...503..721M}, but obtained with less assumptions in the photoionization model with respect to those previous works. The obtained result strongly suggests that HzRGs at $z \sim 3$ are already matured chemically, even in the early Universe where the cosmic age was only $\sim 2$ Gyr.
\item[2.] The inferred physical parameters (gas density, ionization parameter, and gas metallicity) of NLRs in HzRGs show no correlation with the radio power. This suggests that the ionization state of the NLR gas in our sample are not significantly affected by the radio jet.
\end{enumerate}

%%%%%%%%%%%%%%%%%%%%%%%%%%%%
\begin{acknowledgements}
We would like to thank Gary J. Ferland for providing the photoionization code Cloudy to the public.
We also thank Dr. Mitsuru Kokubo for useful comments and suggestions that improved this manuscript.
Photoionization model calculations were in part carried out on the Multi-wavelength Data Analysis System operated by the Astronomy Data Center (ADC), National Astronomical Observatory of Japan. This research has made use of the NASA/IPAC Extragalactic Database (NED), which is operated by the Jet Propulsion Laboratory, California Institute of Technology, under contract with the National Aeronautics and Space Administration.
KT was financially supported by JSPS (19K23452).
TN was financially supported by JSPS (19H00697, 20H01949, and 21H04496).
\end{acknowledgements}
%%%%%%%%%%%%%%%%%%%%%%%%%%%%

\bibliography{radioref}

\end{document}